\documentclass[aps,onecolumn,groupedaddress]{revtex4}
\usepackage{graphics}
\usepackage{amsmath,amssymb}
\begin{document}

%\preprint{}

\title{Torsional fluctuations in columnar DNA assemblies}


\author{D. J. Lee}
%\thanks{}
\affiliation{Department of Chemistry, Faculty of Physical
Sciences, Imperial College London, SW7 2AZ London, UK}

\author{A. Wynveen}
\email[]{awynveen@imperial.ac.uk}
%\thanks{}
\affiliation{Department of Chemistry, Faculty of Physical
Sciences, Imperial College London, SW7 2AZ London, UK}

%\author{A. A. Kornyshev}
%\thanks{}
%\affiliation{Department of Chemistry, Faculty of Physical
%Sciences, Imperial College London, SW7 2AZ London, UK}


\date{\today}

\begin{abstract}
In columnar assemblies of helical bio-molecules the azimuthal degrees of freedom, i.e. rotations about the long axes of molecules, may be important in determining the structure of the assemblies especially when the interaction energy between neighbouring molecules explicitly depends on their relative azimuthal orientations.  For DNA this leads to a rich variety of mesophases for columnar assemblies, each categorized by a specific azimuthal ordering.  In a preceding paper [A.Wynveen, D. J. Lee, and A. A. Kornyshev, Eur. Phys. J. E, 16, 303 (2005)]  a statistical mechanical theory was developed for the assemblies of torsionally rigid molecues in order to determine how thermal fluctuations influence the structure of these mesophases.  Here we extend this theory by including torsional fluctuations of the molecules, where a DNA molecule may twist about its long axis at the cost of torsional elastic energy.  Comparing this with the previous study, we find that inclusion of torsional fluctuations further increases the density at which the transition between the hexagonal structure and the predicted rhombic phase occurs and reduces the level of distortion in the rhombic phase.  As X-ray diffraction may probe the 2-D lattice structure of such assemblies and provide information concerning the underlying interaction between molecules, we have also calculated correlation functions for the azimuthal ordering which are manifest in an x-ray scattering intensity profiles.     
\end{abstract}

%\keywords{}

\maketitle

\section{Introduction}
%\subsection{}
%\subsubsection{}
Condensed DNA exists in a rich variety of phases and mesophases.  
Molecular assemblies of DNA are considered to be lyotropic, so much of 
this rich phase structure may be accessed by changing the concentration of 
DNA in solution.  As the concentration of DNA molecules is increased, the 
orientations of the molecules evolve from a completely disordered isotropic 
phase into a liquid crystal phase. Eventually at large enough densities, 
long range order is established, and a crystalline state is established 
\cite{livolant:96}.  In the liquid crystal phase, various mesophases are 
observed \cite{robinson:61,strey:00,strzelecka:88,durand:92}.  At relatively 
low concentrations the DNA molecules form a cholesteric mesophase, whereas at 
larger concentrations the molecules form columnar assemblies.  We should point 
out that not only is DNA concentration a parameter in determining what 
mesophase the DNA is in, but other factors such as monovalent salt 
concentration, as well as the type and quantity of condensing agent 
used, are equally important \cite{bloomfield:96,podgornik:98}.  
Hence, any analytical study
is necessarily quite involved due to the complexity system.  
\\

Many of these mesophases have been seen in biological systems \cite{reich:94}.
 Determining how these mesophases form is important for understanding how DNA 
packs into viral capsids and sperm heads \cite{bloomfield:96} and may also 
be relevant in the advancement of gene therapies \cite{gelbart:00,strey:98}.  
The study of x-ray diffraction patterns of such mesophases, as well as 
an understanding of the statistical physics underlying DNA assemblies, 
could well elucidate the nature of the forces between such molecules 
\cite{kornyshev:05}.  Furthermore, understanding how properties of these 
mesophases might depend on the sequence of base pair text of the DNA 
that form the mesophase may reveal information concerning the process of 
homologous recognition of genes \cite{weiner:94}.  
\\

Another question that might be probed in the study of DNA mesophases 
is whether the interaction between molecules depends on the azimuthal 
orientations of the DNA about their long axes.  The polyelectrolyte 
model \cite{manning:78,frank:87,levin:02} assumes that the forces between 
molecules are independent of the azimuthal orientations of the molecules.  
This is only valid if the azimuthal dependence of the force is very weak or 
is completely screened by the solvent and its constituent ions.  This 
model, however, is insufficient to describe the various mesophases. 
An alternative model \cite{kornyshev:97a} was proposed, which took into 
account the helical nature of the surface charge patterns on the DNA 
molecule. In this model the interaction energy of a pair of molecules 
depends strongly on the azimuthal orientation out to separations of  
$R>30\AA$.  
This azimuthal dependence was further enhanced when counterions were assumed 
to be preferentially adsorbed within the grooves of the molecule 
\cite{davey}. Such an azimuthal dependence was indeed shown to influence 
both the structures of columnar \cite{harreistot} and cholesteric assemblies
\cite{korntot}.   
\\

In Ref. \cite{harreistot} it was shown that the variety of states observed 
in columnar assemblies are intrinsically linked with the azimuthal 
orientation of the molecules.  These states corresponded to different 
``spin''-orderings, i.e. the configuration of the relative azimuthal 
orientations between neighboring molecules, of the molecules situated on the 
two-dimensional lattice of a columnar assembly.  For example, there 
exist ``ferromagnetic'' and ``antiferromagnetic'' states, i.e. states 
corresponding to where all the molecules are azimuthally aligned and to 
where molecules along a lattice direction have alternating values for 
their azimuthal orientations.  In a later work \cite{wynveen:05a}, 
we extended this ground state calculation \cite{harreistot} to incorporate 
the effects of thermal fluctuations to build up a full statistical 
mechanical model of the columnar assemblies.  Here, two new transitions 
were observed for a fixed hexagonal lattice.  The first corresponded to a 
transition from one of two topologically distinct states of a three-spin 
configuration, where nearest neighbors of a molecule have one of two 
different azimuthal orientations (referred to as the Potts state), to a 
more disordered state where both topologies were equally likely in the 
system. The second was a transition to a Berezinskii-Kosterlitz-Thouless 
like vortex state \cite{bkz}.  Here, also, it was shown that the 
antiferromagnetic state was only stable when the two-dimensional 
hexagonal lattice was distorted.  Allowing for these lattice distortions, 
a phase transition occurring between this rhombic (distorted hexagonal) 
antiferromagnetic state and the ferromagnetic state is observed.  As 
compared to the ground state calculations, incorporating thermal fluctuations 
resulted in a decrease in the mean separation between molecules at 
which this transition occurred, a reduction in the amount the distortion 
to the hexagonal lattice, and a shift from a second order transition 
to a first order one. In both of these studies \cite{harreistot,wynveen:05a}, 
however, the DNA molecules were assumed to be completely rigid.
\\

Torsional flexibility has been considered in a ground state calculation of 
interacting nonhomolgous DNA \cite{cherstvy:04} where the reduction of 
sequence-dependent distortions due to electrostatic interactions between 
the molecules is observed.  This present study, however, considers the 
effects of thermal excitations of the flexible molecules so that torsional 
fluctuations, in which the azimuthal orientation of the double helix is no 
longer uniform along the length of the molecule, occur.  Such effects may be 
implemented relatively easily into an energy functional, which forms a 
starting point for the statistical mechanical treatment.  Previously, 
for rigid molecules, the effect of doubling the length has the same effect as 
halving the temperature.  In flexible molecules, however, length is no longer 
such a trivial parameter in the theory.  Here, shortening the molecules 
plays a similar role to increasing the torsional rigidity since the free 
energy cost of twisting a short molecule to the same degree, i.e. the same 
variation in the azimuthal angle between the ends of the molecule, of that 
of twisting a longer molecule is much greater.  Hence, very short molecules 
may be assumed to be rigid.  Furthermore, as a molecule is lengthened 
beyond a certain torsional persistence length, the extent of the torsional 
fluctuations becomes independent of the length of the molecule. 
\\

Though formulating the energy functional and partition function for columnar assemblies of flexible molecules is a relatively simple task, calculating the free energy and other thermodynamic quantities is not.  Due to the extra degrees of freedom associated with torsional fluctuations, Monte-Carlo methods become less reliable and more time consuming.  Utilizing self-consistent approximations developed through field theoretical methods, however, we encounter a problem arising from treating molecules of finite length; how to account for the freely fluctuating ends of the molecules in the assembly.  Nevertheless, we have managed to incorporate free end effects into the field theoretical calculations and have determined the extent to which they alter the results.  \\

The layout of the paper is as follows.  In Sec. II we consider the case of 
finite temperature interactions between two molecules in parallel 
juxtaposition as a starting point for developing the present calculations 
since the reliability of the these calculations may be tested against the 
``Quantum Mechanical'' formulation of DNA interactions set out in Ref.
\cite{lee:04}.  In Sec. III we summarize the results of the various ground 
state calculations of the system and then incorporate thermal fluctuations at 
the harmonic level (Gaussian fluctuations) for the different configurations 
of the columnar assembly.  Here, we determine the free energy of the assembly 
as well as the correlation function corresponding to the variation of 
relative azimuthal fluctuations between molecules which is important in 
determining the x-ray diffraction patterns of the assemblies 
\cite{kornyshev:05}.  In Sec. IV we go beyond the harmonic treatment of the 
interaction, developing a self-consistent approximation for the assembly.  
This relies on a Hartree approximation for the case where the ends satisfy 
periodic boundary conditions, i.e. both ends of a molecule have the same 
azimuthal orientation, and a correction which takes into account independent 
fluctuations at the ends of each molecule.  At this level of the calculation, 
we demonstrate how torsional fluctuations affect the free energy and 
correlation functions of the assemblies and compare these results to those 
of assemblies composed of rigid molecules \cite{wynveen:05a}.  
We find that the correction accounting for independent fluctuations 
of the ends of the molecules is quite small, and so in first approximation 
may be neglected, demonstrating that the utilization of periodic boundary 
conditions in the calculation is a reasonable approximation.  Finally, 
in Sec. V, we discuss our results and possible future developments.            
\\

\renewcommand{\theequation}{2.\arabic{equation}}
\setcounter{equation}{0}
\section{How to treat molecules of finite length: the 2-body problem}
Our primary aim is to develop a field theoretical framework in which to 
treat the statistical mechanics of flexible molecules condensed in columnar 
assemblies.  As mentioned in the introduction, we first encounter the 
problem of how to treat molecules of finite length.  If we assume that 
periodic boundary conditions can be applied to the ends of the molecules, 
however, we may take advantage of field theoretical techniques that 
already have been developed \cite{periodic}.  By assuming periodic boundary 
conditions we presuppose that both ends of a DNA molecule fluctuate 
torsionally in phase. In other words, both ends share the same 
azimuthal orientation, which isn't necessarily the case. And so within our 
statistical mechanical treatment, we also take into account freely 
fluctuating ends of the molecules.  
\\

Before turning our attention to an assembly, we begin by first considering 
two molecules in parallel.  One of the main motivations for doing this is 
that for the pair interaction, an alternative approach for formulating the 
statistical mechanics of interactions between flexible molecules, which 
resembles that of a quantum mechanical problem, may be employed \cite{lee:04}. 
This, then, provides a check on the reliability of the formulation with which we use to treat the assemblies. 
\\

The  pair potential is a function of the interaxial separation $R$ and the 
relative azimuthal angle $\phi$  between molecules. The relative azimuthal 
angle is defined as the difference in the azimuthal angles between the 
two molecules, $\phi  = \phi _1  - \phi _2 $, where $\phi _i $ 
is the angle that a vector, from the center of the  $i$th molecule to the 
middle of its minor groove, makes with an axis that lies perpendicular to the 
long axes of the molecules. This axis is chosen to pass through the centers of 
both molecules. The pair potential energy per unit length has the form   
\begin{equation}
E_{{\mathop{\rm int}} } (R,\phi ) = \sum\limits_{n = 0} {( - 1)^n a_n (R)} \cos (n\phi ).
\end{equation}
Equation (2.1) is completely model independent and can be deduced purely 
from symmetry requirements. The first requirement is that rotation of one 
of the molecules one whole revolution about its long axis should leave the 
interaction energy unchanged. Hence we may express $E_{{\mathop{\rm int}} }$ 
as a Fourier expansion in terms of the relative azimuthal angle.  Secondly, 
for DNA molecules, helical symmetry dictates that $
 E_{{\mathop{\rm int}} } (R,\phi ) = E_{{\mathop{\rm int}} } (R, - \phi )$, 
so only cosine terms in the Fourier series are retained.  Finally, provided 
that $L \gg H$, where $H$ is the helical pitch of the DNA molecule, helical 
symmetry also ensures that the interaction energy per unit length does not 
depend on $z$, where $z$ is  the coordinate that runs along the long axis 
of the molecule. 
\\

Before proceeding further, we need to consider the behavior of then 
$a_n $coefficients.  We shall use forms for these coefficients obtained 
within the Kornyshev-Leikin (KL) theory \cite{kornyshev:97a} of the 
electrostatic interactions between helical macromolecules.  Such a treatment, 
grounded in Debye-Huckel theory, neglects important non-local effects 
\cite{medvedev:04,kornyshev:99b,solvation} of the dielectric response of 
explicit water. However, such a theory can still provide a qualitative 
picture into the behavior of our system.  Furthermore, this theory should 
work quite well at sufficiently large enough separations between molecules 
and for dilute salt concentrations where Debye-Huckel theory is valid. 
Explicit forms for the $a_n$ coefficients obtained in such a theory are 
given in Ref. \cite{lee:04}.  In the theory, all of the $a_n$ coefficients 
decay exponentially with large interaxial separations with higher order 
terms (larger $n$) decaying at a greater rate.  (The inverse decay lengths 
$\kappa _n$ for each these terms is given by 
$ \kappa _n  = \sqrt {n^2 g^2  + \kappa _s^2 } $, where $ g = 2\pi /H $ 
and $\kappa _s$ is the inverse Debye screening length calculated in Donnan 
equilibrium \cite{harreistot,wynveen:05a,cherstvy:02}.)  Because of this, it 
is sufficient to truncate the series given in Eq. (2.1) at $n=2$ 
\cite{kornyshev:97a}.  The zeroth-order term $a_0$, which describes 
the interaction between two cylinders with a uniform charge distribution, 
does not depend on $\phi$ and therefore need not be considered in the 
treatment of torsional fluctuations.  However, it does play an important 
role in the positional structure 
of the assembly.  Finally, since $a_2$ decays 
faster than $a_1$, minimizing the interaction energy with respect to 
$\phi$ leads to a configuration  below a critical value $R_*$ of the 
interaxial spacing where the preferred value of $\phi$ is non-zero, 
whereas above $R_*$ it is zero.  Even when including non-local effects, 
these qualitative features should still be manifest in the interactions.
\\ 

Upon including thermal fluctuations of the relative twisting of one DNA 
molecule with respect to another, $\phi$ now must be assigned a 
$z$-dependence.  Following from previous work \cite{lee:04}, we write down 
the partition function as a functional or path integral
\begin{equation}
\int {\cal D} \phi (z) \exp \left( { - \frac{1}{{k_B T}}\int_{ - L/2}^{L/2} {dz\left[ {\frac{C}{4}\left( {\frac{{d\phi }}{{dz}}} \right)^2  - a_1 \cos (\phi ) + a_2 \cos (2\phi )} \right]} } \right),
\end{equation}
where $C$ is defined as the torsional elasticity modulus of the helices.  
Here, we have assumed that the centers of both molecules are at $z=0$
and each molecule has length $L$.
The free energy then can be found upon calculating the partition function.
\\

As was discussed previously in Ref. \cite{lee:04}, with this partition 
function the problem can be recast in an alternative formulation that 
mirrors a quantum mechanical problem.  Here, we solve the Schr\"{o}dinger 
equation
\begin{equation}
- \frac{1}{2}\frac{{d^2 \psi _E (\phi )}}{{d\phi ^2 }} + V(\phi )\psi _E (\phi ) = E\psi _E (\phi ),
\end{equation}       
where
\begin{equation}
V(\phi ) =  - \frac{{\lambda _p^2 }}{{\lambda _0^2 }}\cos (\phi ) + \frac{{a_2 \lambda _p^2 }}{{a_1 \lambda _0^2 }}\cos (2\phi ),
\end{equation}
$ \lambda _0^2  = C/(2a_1 )$ and $ \lambda _p  = C/(2k_B T)$. In this formulation, the free energy is expressed through the solutions of Eq. (2.3) as 
\begin{equation}
F =  - k_B T\ln \sum\limits_E {\psi _E^ *  (\phi _ +  )\exp \left( { - \frac{{EL}}{{\lambda _p }}} \right)\psi _E (\phi _ -  ) - k_B T\ln \Theta }, 
\end{equation}
where $\Theta$ is a constant which can be neglected when comparing 
free energies of different states.  $\phi_+$ and $\phi_-$ are the 
values of $\phi$ at the ends, $z=L/2$ and $z=-L/2$   respectively, of the 
molecules.  Periodic boundary conditions amount to setting $\phi_+ = \phi_-$, 
and computing the free energy entails summing over $\phi_+$. However, 
as was pointed out before, both ends should be left to fluctuate freely.  
Consequently, we should integrate over both $\phi_-$ and $\phi_+$, 
allowing $\phi_-$
to take on any value between $0$ and $2 \pi$, and $\phi_+$ to take on any 
value whatsoever.  When the fluctuations are small we may expand 
$V(\phi )$ to quadratic order in $\phi$ around its preferred value in the 
ground state.  We shall consider only the case when the preferred value is 
zero (large interaxial spacings), although this calculation may be easily 
extended to a non-zero preferred angle.  Eq. (2.3) may be solved analytically 
to determine the free energy.  Some of the details of the calculation of $F$
are given in Appendix A.  Here, we shall quote the end result for long 
molecules 
 \begin{equation}
F \simeq  - k_B T \Theta{^{'}}  + E_0  + {\frac{CL}{4 \lambda \lambda _p }} 
+ \frac{{k_B T}}{2}\ln \left( {\frac{{\lambda _p }}{\lambda }} \right) 
- \frac{{k_B T}}{2}\exp \left( { - \frac{{2L}}{\lambda }} \right),
\end{equation}
where $ E_0  =  - a_1  + a_2 $ and $ \lambda ^2  = C/(2(a_1  - 4a_2 )) $. The last two terms are leading order corrections due to the finite length of the molecules. 
\\

Returning to Eq. (2.2), we now approach the problem directly through path integration, first in the Gaussian or harmonic approximation. In such an approximation, both cosine terms are expanded out to quadratic order in $\phi$ around the preferred orientation.  This is equivalent to expanding $V(\phi)$
to quadratic order in $\phi$ in the previous formulation.  We have demonstrated this equivalence in Ref. \cite{lee:04}.  In this approximation the partition 
function becomes
\begin{equation}
Z = \exp \left( { - \frac{{E_0 }}{{k_B T}}} \right)\int {\cal D} \phi {\rm  }\exp \left( { - \frac{1}{{k_B T}}\int_{ - L/2}^{L/2} {dz\left[ {\frac{C}{4}\left( {\frac{{d\phi }}{{dz}}} \right)^2  + \frac{m}{2}\phi ^2 } \right]} } \right),
\end{equation}
where $ m = a_1  - 4a_2 $. We shall consider the following ansatz for the 
form of $\phi (z) $;
\begin{equation}
\phi (z) = \phi _p (z) + \frac{{\gamma z}}{L} + \phi _0, 
\end{equation}
where $ \phi _p (z) $ is the component of $ \phi (z)$ that satisfies 
periodic boundary conditions, i.e. 
$\phi _p ( - L/2) = \phi _p (L/2)$ and it has a spatial average of 
zero along the length of the molecules.
\\

Having a period of $L$, $ \phi _p (z)$ may be expressed as 
\begin{equation}
\phi _p (z) = \frac{1}{{\sqrt L }}\sum\limits_{n \ne 0} {b_n \exp \left( {\frac{{2\pi inz}}{L}} \right)}. 
\end{equation}
The $\phi_0$ component of $ \phi (z)$ is used for the calculation of rigid body fluctuations of the average relative azimuthal orientation between the two molecules and so does not depend on $z$. The component of $ \phi (z) $ proportional to $\gamma$ is included in order to account for free fluctuations of the ends by allowing $\gamma$ to vary, where $\gamma$ is the difference in the azimuthal angles between the ends of the molecules. We may recast Eq. (2.7) in the following way, using Eqs. (2.8) and (2.9),
\begin{eqnarray}
&&  Z = \exp \left( { - \frac{{E_0 }}{{k_B T}}} \right)\int {d\phi _0 } 
\int {d\gamma } \prod\limits_{n \ne 0} {\int d } b_n {\rm  } \times 
\nonumber \\
&& \exp \left( { - \frac{1}{{k_B T}}\sum\limits_{n \ne 0} {\left[ {\frac{1}{{2G_n }}b_n b_{ - n}  + \frac{{\gamma m}}{2}J_n b_{ - n}  + \frac{{\gamma m}}{2}J_{ - n} b_n } \right] - \frac{1}{{k_B T}}\left[ {\frac{C}{{4L}} + \frac{{Lm}}{{24}}} \right]\gamma ^2 }  + \frac{m}{{2k_B T}}\phi _0^2 } \right),
\end{eqnarray}
where $ G_n  = \left( {\frac{C}{2}\left( {\frac{{2\pi n}}{L}} \right)^2  + m} \right)^{ - 1} $ and $
J_n  = \frac{1}{{\sqrt L }}\int\limits_{ - L/2}^{L/2} {dz\left( {\frac{z}{L}} \right)\exp \left( { - \frac{{2\pi inz}}{L}} \right)}  = \frac{{i( - 1)^n (1 - \delta _{n,0} )}}{{(2\pi n)}} $.
\\

The crucial step is to make the variable shift 
$ b_n  \to b_n  - m\gamma J_n G_n $ to decouple $b_n$ from $\gamma$. 
In the domain $ - L/2 \le z < L/2$, $ \frac{{\gamma z}}{L}$   may be written 
as a Fourier series and the effect of the variable shift is to adsorb this 
into $\phi _p (z)$. At $z = L/2$, however,
$ \frac{{\gamma z}}{L} $ cannot be expressed as a Fourier series.  At this 
point there is a difference of $\gamma$ in the sum of the Fourier series 
representing $\frac{{\gamma z}}{L}$ and the actual value of 
$\frac{{\gamma z}}{L}$. The torsional energy term, however, contains 
derivatives of $\phi$  which are sensitive to this difference.  
Therefore, when we make this shift we find that the integrand in our 
partition function still depends on $\gamma$. Thus this leads to a difference 
in the partition function for free boundary conditions as opposed to that 
assuming periodic boundary conditions \cite{flucends}.\\

The partition function thus becomes
\begin{equation}
Z = Z_0 Z_f Z_p \exp \left( { - \frac{{E_0 }}{{k_b T}}} \right),
\end{equation}
where $
Z_0  = \int_{ - \infty }^\infty  {d\phi _0 } \exp \left( { - \frac{m}{{2k_B T}}\phi _0^2 } \right) $ 
corresponds to the component of the partition function that depends only on the average azimuthal orientation of the molecules, $
Z_f  = \int\limits_{ - \infty }^\infty  {d\gamma } \exp \left( { - \frac{{\gamma ^2 \lambda _p }}{{4\lambda }}\coth \left( {\frac{L}{{2\lambda }}} \right)} \right) $ 
is the part of the partition function that takes into account the free rotations of the ends of the molecules , and 
$ Z_p  = \prod\limits_{n \ne 0} {\int\limits_{ - \infty }^\infty  {db_n } } \exp \left( { - \frac{1}{{2k_B T}}\sum\limits_{n \ne 0} {\frac{{b_n b_{ - n} }}{{G{}_n}}} } \right) $ is the component that takes into account all $z$-dependent azimuthal angle orientations in the case where the ends of the molecules share the same orientation.\\

From Eq. (2.11) we may compute the free energy (Appendix B). We find that
\begin{equation}
F \simeq  - k_B T\Theta ^{'}  + E_0  + \frac{{CL}}{{4\lambda \lambda _p }} + \frac{{k_B T}}{2}\ln \left( {\frac{{\lambda _p }}{\lambda }} \right) + \frac{{k_B T}}{2}\ln \left( {1 - \exp \left( { - \frac{{2L}}{\lambda }} \right)} \right).
\end{equation}
Expanding the log of the last term when $L$ is large, we retrieve Eq. (2.6).  
And so we find that this formulation is consistent with the exact one of 
Ref. \cite{lee:04}.  We therefore are ready to tackle the many body 
problem of assemblies.
\\

%%%%%%%%%%%%%%%%%%%%%%%%%%%%%%%%%%%%%%%%%%%%%%%%%%%%%%%%%%%%%%%%%%%%

\renewcommand{\theequation}{3.\arabic{equation}}
\setcounter{equation}{0}
\section{Assemblies of molecules of finite length: the Gaussian approximation}

For an assembly, the energy can be written as a sum of the energies of each molecular pair over the entire lattice that defines the assembly as 
\begin{eqnarray}
E[\phi ] = \int\limits_{ - L/2}^{L/2} {dz} \sum\limits_{j,l} {\left[ {\frac{C}{2}\left( {\frac{{d\phi _{jl} (z)}}{{dz}}} \right)^2  + \sum\limits_{n = 0} {( - 1)^n a_n (R_2 )} \cos \left( {n\left( {\phi _{jl} (z) - \phi _{j + 1l - 1} (z)} \right)} \right)} \right.} \nonumber \\
\left. { + ( - 1)^n a_n (R_1 )\left( {\cos \left( {n\left( {\phi _{jl} (z) - \phi _{j - 1l} (z)} \right)} \right) + \cos \left( {n\left( {\phi _{jl} (z) - \phi _{jl - 1} (z)} \right)} \right)} \right)} \right].
\end{eqnarray}
Here, we have introduced two lattice vectors $ \vec u_i  = jR_1 \hat u$ and $ \vec v_j  = lR_1 \hat v
$, where $\hat u$ and $\hat v$ are unit vectors that describe the relative positions between molecules situated at the sites of the two dimensional lattice. These two vectors, as well as the convention used in labeling the lattice sites, are shown in Fig.1. 
\\

%%%%%%%%%
\begin{figure}
\includegraphics[12cm,22cm][13cm,23cm]{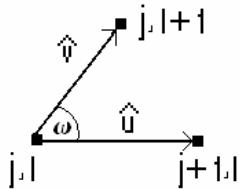}
\vspace{2.5cm} \caption{Lattice labeling and relative positions of  
the lattice vectors $
\hat u
$
 and $
\hat v$ .}
\end{figure}
%%%%%%%%%%%%%%%%%%%%%%%%%%%%%%%%%%%%%%%%%%%%%%%%%%%%%%%%%%%%%%

%%%%%%%%%%%%%%%%%%%%%%%%%%%%%%%%%%%%%%%%%%%%%%%%%%%%%%%%%%%%%%%%%%%%%%

In Eq. (3.1) we have included only the interactions between nearest 
neighbors and have assumed that all the DNA molecules are of the same 
length with the center of each molecule lying at $z=0$ \cite{azde}.   
Again we may truncate the series at $n=2$. In calculating the energy of the 
lattice, we have allowed for two separations,$R_1$ and $R_2$, of the six 
nearest neighbors of a given molecule, since there may be distortions 
from the hexagonal lattice \cite{harreistot,wynveen:05a}.  For this 
distorted or ``rhombic'' lattice, $R_1$ corresponds to the distance to 
the neighboring molecule at an adjacent corner of the unit cell 
while $R_2$ is the distance across the short diagonal of the rhombic 
cell. We may define a distortion angle $\omega$ (shown in Fig. 1) that 
characterizes the relationship between the two separations in Eq. (3.1), 
$ R_2  = R_1 \sqrt {2 - 2\cos \omega } $.  In the ground state, the amount 
of distortion is determined by minimization of $E[\phi ]$ with respect to 
the distortion angle $\omega$. The relative strengths of the $a$ 
coefficients determine the degree of this distortion, e.g. the $a_0$ term, which contributes a repulsive component to the overall force, has the effect of reducing the amount of rhombic distortion as on its own it favors a 
hexagonal state.  For certain azimuthal configurations of the molecules 
at large densities, the distortion angle $\omega$ is greater than $60^{\rm o}$.
\\

Before considering effects of thermal fluctuations, let us first describe 
the various ground states that the assembly has.  When molecular separations 
are large, all the relative azimuthal orientations of the molecules are 
the same.  This state may be referred to as the ``ferromagnetic'' state.  
For this state the lattice is hexagonal, $R_1  = R_2$. Below a critical 
value of the separation of nearest neighbors, $R_*^1$, the assembly adopts 
a new state where four of the six nearest neighbors about a specific 
molecule adopt  a different azimuthal orientation to that molecule, 
as shown in Fig. 2.  This state then may be termed the ``antiferromagnetic'' 
state.  This configuration of the azimuthal orientations of the molecules 
favors rhombic distortions, $R_1  \ne R_2$, described in the previous 
paragraph.  The ground state azimuthal configuration for this configuration 
is characterized by the following equations
\addtocounter{equation}{+1}
\begin{align*}
\left( {\phi _{jl} (z) - \phi _{j + 1l - 1} (z)} \right) & = 0 &  
{\text{(across the short diagonal of the rhombic unit cell)}}&
\tag{{\theequation}a}
\end{align*}
\begin{align*}
\left( {\phi _{jl} (z) - \phi _{j - 1l} (z)} \right) = 
\left( {\phi _{jl} (z) - \phi _{jl - 1} (z)} \right) & = \psi 
&{\text{(adjacent corners of the unit cell)}}&
\tag{{\theequation}b}
\end{align*}
where $ \cos (\psi ) = \frac{{a_1 (R_1 )}}{{4a_2 (R_1 )}} $. For even 
denser assemblies, a new critical value of average separation $R_*^2$ is 
realized where the most energetically favorable state is the 
``Potts'' state.  In this state, the lattice returns to a hexagonal confirmation.  As shown in Fig. 2, the azimuthal orientations of the molecules for this state may have one of three different values: $\phi_0$,
$\phi _0  + \psi _p$, and $\phi _0  + 2 \psi _p$ where
\begin{equation}
\cos \left( {\psi _p } \right) = \frac{1}{4}\left( {1 + \sqrt {1 + \frac{{2a_1 }}{{a_2 }}} } \right).
\end{equation}
\\

\begin{figure}
\includegraphics[5cm,24.5cm][6cm,25.5cm]{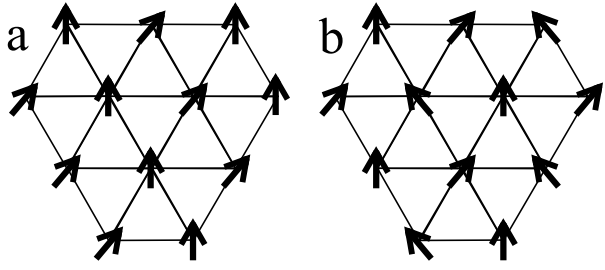}
\vspace{2.5cm} \caption{Schematic pictures of the azimuthal orientations 
of molecules in two of the ground state configurations.  The first 
(a) corresponds to the antiferromagnetic state where the molecules adopt 
layers in which the molecules are in the same azimuthal alignment. 
The  second (b) corresponds to the Potts state, where the azimuthal 
orienations of the molecules have one of three values as defined 
in the text.}
\end{figure}

Upon introducing thermal fluctuations, we shall only consider the 
ferromagnetic and antiferromagnetic states.  (For a discussion of the Potts 
state, which, again, only occurs at very large densities, at finite 
temperatures for rigid molecules see Ref. \cite{wynveen:05a}.)  First, for 
the antiferromagnetic state, the partition function in the 
Gaussian approximation may be written as  
\begin{equation}
Z = \exp \left( { - \frac{{E_0^{AF} }}{{k_B T}}} \right)\prod\limits_{jl} {\int {\cal D} \phi '(z)} {\rm  }\exp \left( { - \frac{{E_H^{AF} [\phi ']}}{{k_B T}}} \right),
\end{equation}
where
\begin{equation}
E_H^{AF} [\phi '] = \int\limits_{ - L/2}^{L/2} {dz} \sum\limits_{j,l} {\left[ {\frac{C}{2}\left( {\frac{{d\phi '_{jl} (z)}}{{dz}}} \right)^2  + \frac{{m_2 }}{2}\left[ {\phi '_{jl} (z) - \phi '_{j + 1l - 1} (z)} \right]^2  + \frac{{m_1 }}{2}\left[ {\left( {\phi '_{jl} (z) - \phi '_{j - 1l} (z)} \right)^2  + \left( {\phi '_{jl} (z) - \phi '_{jl - 1} (z)} \right)^2 } \right]} \right]}, 
\end{equation}
and the ground state energy is $
E_0^{AF}  = a_0 (R_2 ) - a_1 (R_2 ) + a_2 (R_2 ) + 2a_0 (R_1 ) - 2a_1 (R_1 )\cos (\psi ) + 2a_2 (R_1 )\cos (2\psi )$. 
The azimuthal orientations of the molecules at the lattice points are replaced with the following values: 
\begin{eqnarray}
\phi _{jl} (z) - \phi _{j - 1l} (z) = \psi  + \phi '_{jl} (z) - \phi '_{j - 1l} (z)
\nonumber \\
\phi _{jl} (z) - \phi _{jl - 1} (z) = \psi  + \phi '_{jl} (z) - \phi '_{jl - 1} (z)
 \\
\phi _{jl} (z) - \phi _{j + 1l - 1} (z) = \phi '_{jl} (z) - \phi '_{j + 1l - 1} (z). \nonumber 
\end{eqnarray}
Here, $ m_2  = a_1 (R_2 ) - 4a_2 (R_2 )$ and 
$ m_2  = a_1 (R_1 )\cos (\psi ) - 4a_2 (R_1 )\cos (2\psi )
$. As in the previous section, we introduce the following ansatz,
\begin{equation}
\phi '_{jl} (z) = \phi _{jl}^p (z) + \frac{{\gamma _{jl} z}}{L},
\end{equation}
splitting up the azimuthal value along each molecule at each lattice site 
into a function $\phi _{jl}^p (z) $ that satisfies the periodic boundary 
conditions, $ \phi _{jl}^p ( - L/2) = \phi _{jl}^p (L/2)
$, linking the ends of the molecules and the correction 
$ \frac{{\gamma _{jl} z}}{L}$ to account for independent fluctuations of 
the two ends of each molecule. We may express $\phi _{jl}^p (z)$ in a 
similar form as Eq. (2.9) for the previous section (see Appendix C). The 
rigid body part of $\phi _{jl} (z) $ has been included in 
$\phi _{jl}^p (z)$ and is the spatial average of $\phi _{jl}^p (z)$ along 
the molecule.  The steps in calculating the free energy in the Gaussian 
approximation are similar to those given in the previous section 
(see Appendix C) with the end result being
\begin{eqnarray}
F = \frac{{k_B T}}{{2(2\pi )^2 }}\int\limits_{ - \pi }^\pi  {dx} \int\limits_{ - \pi }^\pi  {dy\ln \left( {\frac{{\lambda _p }}{{\lambda _1 }}\left( {\frac{{\hat C(x,y;\alpha )}}{{\hat C(x,y;1)}}} \right)^{1/2} \coth \left( {\frac{{\hat C(x,y;\alpha )^{1/2} L}}{{2\lambda _1 }}} \right)} \right)} 
\nonumber \\
+ \frac{{k_B T}}{{(2\pi )^2 }}\int\limits_{ - \pi }^\pi  {dx} \int\limits_{ - \pi }^\pi  {dy} \ln \left( {\frac{1}{{\hat C(x,y,1)}}\sinh \left( {\frac{L}{{2\lambda _1 }}\hat C(x,y,\alpha )} \right)} \right) + k_B T\Theta _{asb}, 
\end{eqnarray}
where $ \lambda _1  = \sqrt {\frac{C}{{2\left( {a_1 (R_1 )\cos (\psi ) - 4a_2 (R_1 )\cos (2\psi )} \right)}}} $ , $ \lambda _2  = \sqrt {\frac{C}{{2\left( {a_1 (R_2 ) - 4a_2 (R_2 )} \right)}}} 
$, $ \hat C(x,y;\alpha ) = (1 - \cos (x)) + (1 - \cos (y)) + \alpha (1 - \cos (x - y))
$ and $\alpha  = \left( {\lambda _1 /\lambda _2 } \right)^2$.  
For the ferromagnetic state we simply set
$\psi=0$ and $R_1  = R_2$. 
\\

Additionally, we may compute the correlation functions associated with the 
thermally induced fluctuations of the azimutathal orientation of the 
molecules.  Calculation of these are particularly useful since they are 
reflected in the intensity profiles of x-ray diffraction patterns of 
columnar assemblies \cite{kornyshev:05}.  This azimuthal correlation 
function within the Gaussian approximation is defined as 
\begin{equation}
\left\langle {\exp \left( {in\left( {\phi '_{jl} (z) - \phi '_{j'l'} (z')} \right)} \right)} \right\rangle _0  = \exp \left( { - \frac{{E_0^{AF} }}{{k_B T}}} \right){\rm  }\frac{1}{Z}\prod\limits_{jl} {\int {\cal D} \phi '(z)} \exp \left( {in\left( {\phi '_{jl} (z) - \phi '_{j'l'} (z')} \right)} \right)\exp \left( { - \frac{{E_H^{AF} [\phi ']}}{{k_B T}}} \right).
\end{equation}
This may be written as (summarized in Appendix D) 
\begin{eqnarray}
\left\langle {\exp \left( {in\left( {\phi '_{jl} (z) - \phi '_{j'l'} (z')} \right)} \right)} \right\rangle _0  & = & \exp \left( {n^2 \left( {\left\langle {\phi '_{jl} (z)\phi '_{j'l'} (z')} \right\rangle _0  - \frac{1}{2}\left\langle {\phi '_{jl} (z)^2 } \right\rangle _0  - \frac{1}{2}\left\langle {\phi '_{jl} (z')^2 } \right\rangle _0 } \right)} \right) \nonumber \\
 & \equiv & \exp \left( {n^2 G(j - j',l - l',z,z')} \right).
\end{eqnarray}
$G(j - j',l - l',z,z')$ may be split up into terms corresponding to three contributions 
\begin{equation}
G(s,r,z,z') = G_0 (s,r) + G_P (s,r,z - z') + G_{AP} (s,r,z,z'),
\end{equation}
where, $ s = j - j'$ and $ r=l-l'$.
\\

The first contribution takes into account rigid body fluctuations of the azimuthal  orientation of the DNA molecules.  It has the following form for large separations, i.e. $ s,r \gg 1 $ (for a general expression see Appendix D)
\begin{equation}
G_0 (r,s) \approx  - \frac{{k_B T\lambda _1^2 }}{{\pi LC\sqrt {1 + 2\alpha } }}\left( {\ln \left( {r^2  + s^2  + \frac{{2\alpha rs}}{{(1 + \alpha )}}} \right)} \right) - \Delta _\psi ^2, 
\end{equation}
where $ \Delta _\psi ^2  = G_0 (1,0) = G_0 (0,1) = \frac{{2k_B T}}{{LC}}\frac{{\lambda _1^2 }}{\pi }\arcsin \left( {\frac{1}{{\sqrt {2(\alpha  + 1)} }}} \right) $.
The first term in Eq. (3.12) grows with increasing separation because of there is no long range order for such a 2-D system. The second contribution of Eq. (3.9) represents the torsional (z-dependent) azimuthal fluctuations assuming periodic boundary conditions. Again, when the separations are large,
\begin{eqnarray}
 &G_P (j - j',l - l',z - z') \approx  - \Delta _\phi ^2   \nonumber \\ 
  &=  - \frac{{k_B T}}{{2C}}\frac{1}{{(2\pi )^2 }}\int\limits_{ - \pi }^\pi  {dx} \int\limits_{ - \pi }^\pi  {dy} \left( {\frac{{\lambda _1 }}{{\hat C(x,y;\alpha )^{1/2} }}\coth \left( {\frac{{L\hat C(x,y;\alpha )^{1/2} }}{{2\lambda _1 }}} \right) - \frac{{2\lambda _1^2 }}{{L\hat C(x,y;\alpha )}}} \right).  
\end{eqnarray}
The last contribution is the correction due to allowing both ends of each molecule to fluctuate independently. This takes the form for large separations of
\begin{eqnarray}
G_{AP}^\infty  (z,z') = \frac{{k_B T\lambda _1 }}{4C}\frac{1}{{(2\pi )^2 }}\int\limits_{ - \pi }^\pi  {dx} \int\limits_{ - \pi }^\pi  {dy} \frac{1}{{\hat C(x,y;\alpha )^{1/2} }}\exp \left( { - \frac{{L\hat C(x,y;\alpha )^{1/2} }}{{\lambda _1 }}} \right)\left( {1 - \exp \left( { - \frac{{2L\hat C(x,y;\alpha )^{1/2} }}{{\lambda _1 }}} \right)} \right)^{ - 1} \nonumber \\
\times \left[ \left( {\exp \left( {\frac{{\left| z \right|C(x,y;\alpha )^{1/2} }}{{\lambda _1 }}} \right) - \exp \left( {\frac{{ - \left| z \right|C(x,y;\alpha )^{1/2} }}{{\lambda _1 }}} \right)} \right)^2 \right. \\ 
\left.
+ \left( {\exp \left( {\frac{{\left| {z'} \right|C(x,y;\alpha )^{1/2} }}{{\lambda _1 }}} \right) - \exp \left( {\frac{{ - \left| {z'} \right|C(x,y;\alpha )^{1/2} }}{{\lambda _1 }}} \right)} \right)^2  \right]. \nonumber
\end{eqnarray}
And so the correlation function at large separations can be written as 
\begin{eqnarray}
 \left\langle {\exp \left( {in\left( {\phi '_{jl} (z) - \phi '_{j'l'} (z')} \right)} \right)} \right\rangle _0  \approx \left( {(j - j')^2  + (l - l')^2  + \frac{{2\alpha (l - l')(j - j')}}{{(1 + \alpha )}}} \right)^{ - \gamma }   \nonumber \\ 
 {\rm                                               } \times \exp \left( { - n^2 \left( {\Delta _\phi ^2  + \Delta _\psi ^2 } \right)} \right)\exp \left( { - n^2 G_{AP}^\infty  (z,z')} \right), 
 \end{eqnarray}
where $ \gamma  = n^2 k_B T\lambda _1^2 /(\pi LC\sqrt {1 + 2\alpha } ) $. As the separations increase, the azimuthal correlation is reduced as one might expect. This, again, reflects the fact that in such a 2-D system there is no long-range order.
\\

%%%%%%%%%%%%%%%%%%%%%%%%%%%%%%%%%%%%%%%%%%%%%%%%%%%%%%%%%%%%%%%%%%%%%%%%

\renewcommand{\theequation}{4.\arabic{equation}}
\setcounter{equation}{0}
\section{Assemblies of molecules of finite length: the 
self-consistent approximation.} 

Near the point of frustration, i.e. the location of the transition between 
the distorted ``antiferromagnetic'' state and the hexagonal ``ferromagnetic''
 state, the fluctuations become quite large so that the Gaussian 
approximation is no longer valid.  Thus we must extend the calculation by 
using a self-consistent approximation, namely, a Hartree approximation.  
Unfortunately, we were unable to find a Hartree approximation that takes 
into account freely fluctuating ends of the molecules, but it is possible to 
treat this contribution as a small correction to the Hartree approximation 
for periodic boundary conditions.  In such an approximation it is possible 
to determine how torsionally softening or changing a molecule's 
length alters the value of $R_*^1 (T)$, the separation at which the phase 
transition occurs.  Furthermore, the correlation functions of the previous 
section also can be calculated within this approximation.
\\

The self-consistent approximation first entails carrying out a series 
expansion of Eq. (3.1) in powers of $\phi '$ where anharmonic terms, 
terms of $O(\phi '^3 )$ or greater, are treated as perturbations to 
the Gaussian approximation.  The value of $\psi$ is determined 
through the requirement that  
\begin{equation}
\left\langle {\phi '_{jl} (z) - \phi '_{j - 1l} (z)} \right\rangle  = \left\langle {\phi '_{jl} (z) - \phi '_{jl - 1} (z)} \right\rangle  = \left\langle {\phi '_{jl} (z) - \phi '_{j + 1l - 1} (z)} \right\rangle  = 0.
\end{equation}
(Details of such a perturbation expansion are given in Appendix E) 
Terms for the correlation function and the free energy, where periodic 
boundary conditions are assumed, may be re-summed in a similar fashion to 
the Hartree approximation calculations of Refs. \cite{wynveen:05a,lee:04}.  
Following similar steps (given in Appendix F), we obtain the following 
Hartree result for the free energy per molecule  
\begin{eqnarray}
F_H  = \frac{{k_B T}}{{(2\pi )^2 }}\int\limits_{ - \pi }^\pi  {dx} \int\limits_{ - \pi }^\pi  {dy\ln \left( {\sinh \left( {\frac{{L\hat C(x,y;\alpha _H )^{1/2} }}{{2\lambda _1^H }}} \right)} \right) - 2La_1 (R_1 )\cos \psi _H \exp \left( { - \frac{{\lambda _1^H }}{{4\lambda _p }}\chi _1 \left( {\frac{L}{{2\lambda _1^H }},\alpha _H } \right)} \right)} \nonumber \\ 
+ 2La_2 (R_1 )\cos 2\psi _H \exp \left( { - \frac{{\lambda _1^H }}{{\lambda _p }}\chi _1 \left( {\frac{L}{{2\lambda _1^H }},\alpha _H } \right)} \right) - La_1 (R_2 )\exp \left( { - \frac{{\lambda _1^H }}{{4\lambda _p }}\chi _2 \left( {\frac{L}{{2\lambda _1^H }},\alpha _H } \right)} \right)
\\
+ La_2 (R_2 )\exp \left( { - \frac{{\lambda _1^H }}{{\lambda _p }}\chi _2 \left( {\frac{L}{{2\lambda _1^H }},\alpha _H } \right)} \right) - \frac{{LC}}{{8\lambda _1^H \lambda _p }}\left( {2\chi _1 \left( {\frac{L}{{2\lambda _1^H }},\alpha _H } \right) + \alpha _H \chi _2 \left( {\frac{L}{{2\lambda _1^H }},\alpha _H } \right)} \right),
\nonumber 
\end{eqnarray}
where $\alpha _H  = \left( {\lambda _1^H /\lambda _2^H } \right)^2$, and $\lambda _1^H$ and $\lambda _2^H$ satisfy the following transcendental equations 
 \begin{eqnarray}
\lambda _1^H  = \sqrt {\frac{C}{{2\left( {a_1 (R_1 )\cos (\psi _H )\exp \left( { - \frac{{\lambda _1^H }}{{4\lambda _p }}\chi _1 \left( {\frac{L}{{2\lambda _1^H }},\alpha _H } \right)} \right) - 4a_2 (R_1 )\cos (2\psi _H )\exp \left( { - \frac{{\lambda _1^H }}{{\lambda _p }}\chi _1 \left( {\frac{L}{{2\lambda _1^H }},\alpha _H } \right)} \right)} \right)}}} \nonumber \\
\lambda _2^H  = \sqrt {\frac{C}{{2\left( {a_1 (R_2 )\exp \left( { - \frac{{\lambda _1^H }}{{4\lambda _p }}\chi _2 \left( {\frac{L}{{2\lambda _1^H }},\alpha _H } \right)} \right) - 4a_2 (R_2 )\exp \left( { - \frac{{\lambda _1^H }}{{\lambda _p }}\chi _2 \left( {\frac{L}{{2\lambda _1^H }},\alpha _H } \right)} \right)} \right)}}}. 
\end{eqnarray}
$\psi_H$ satisfies the following relation
\begin{equation}
\cos \left( {\psi _R^H } \right) = \frac{{a_1 }}{{4a_2 }}\exp \left( {\frac{{3\lambda _1^H }}{{4\lambda _p }}\chi _1 \left( {\frac{L}{{2\lambda _1^H }},\alpha _H } \right)} \right).
\end{equation}
The functions $ \chi _1 \left( {\frac{L}{{2\lambda _1^H }},\alpha _H } \right) $ and $
\chi _2 \left( {\frac{L}{{2\lambda _1^H }},\alpha _H } \right) $ 
are given in Appendix E.  The free energy is then minimized with respect 
to $\omega$  to determine the degree of distortion of the hexagonal 
lattice for the antiferromagnetic state. The ferromagnetic state is simply 
obtained by setting 
$\psi=0$ so that $R_1=R_2$ and, thus, 
$ \lambda _1^H  = \lambda _2^H  = \lambda _H $ and $\alpha=1$.
\\

It is quite interesting to look at the Hartree result for the ferromagnetic 
state for infinitely long molecules, which is very similar to the result 
given in Ref. \cite{lee:04} for a pair of molecules, since it provides 
some insight into the nature of the many body effects of the assembly.  
There are two differences:  first of all, $ \lambda _p $ is 
replaced by $\lambda _p C_{mb}^{ - 1}$, where $C_{mb}$ 
is a constant resulting from the many body effects of the assembly; and 
secondly, an overall factor of three that multiplies the free energy is 
included to account for the coordination number of nearest neighbors 
about a molecule.  We calculate that $ {\cal C}_{mb}  \simeq 0.560$ so  
we find many body effects suppress thermal fluctuations.  This is not 
surprising considering that the many neighbors in the assembly should 
increase the effective interaction between molecules.  
Therefore, thermal fluctuations of the same magnitude in the assembly 
cost more energy than that for a single pair of molecules. 
\\

We may also show that the results above conform to that of the formulation 
of rigid molecules \cite{wynveen:05a}.  Changing variables to 
$ J_1^H $ and $J_2^H $ where $ \lambda _1^H  = \sqrt {CL/(2J_1^H )}$ 
and $\lambda _2^H  = \sqrt {CL/(2J_2^H )}$, we may rewrite Eq. (4.3) in 
terms of $J_1^H$ and $J_2^H$. Upon taking the limit $ L \to 0 $, it is 
fairly straightforward to recover the Hartree results for the 
configurational states of Ref. \cite{wynveen:05a}.  Taking the limit   
of these equations, i .e. the molecules are treated as completely rigid, 
also yields this same result.  This is fully consistent with the physics 
we might expect.  As we shorten the length of the molecule, the energy 
cost of torsional fluctuations begins to counter the reduction in free 
energy due to the entropy gain associated with them.  And so with decreasing 
length, these fluctuations become suppressed and rigid body fluctuations 
start to dominate. 
\\

We first examine how the phase boundary   between the ferromagnetic 
hexagonal state and antiferromagnetic rhomic state changes with the length 
and flexibilty of the molecules assuming periodic boundary conditions.  
Equating the free energy of the antiferromagnetic state with that of the 
ferromagnetic state, Eqs. (4.2)-(4.4), we can determine the location of the 
transition between the two states. The location of this transition in terms 
of the average interaxial spacing between molecules in the assembly as a 
function of molecular length is shown in Fig. 3 for molecules with varying 
torsional flexibilities.  Increasing the flexibility of the molecule 
results in the transition moving to larger densities (smaller interaxial 
spacings) as one might expect.  With increasing flexibility, there is an 
increase in thermally induced torsional fluctuations and so the location 
of the transition as the flexibility is increased moves in the same direction 
as when the molecule is shortened, where thermally induced rigid-body 
fluctuations become larger.  Also shown, in Fig. 4, is the level of 
distortion (value of the rhombic angle  ) from the hexagonal lattice as 
a function of the average interaxial spacing for the antiferromagnetic 
state.  As seen in the plot, greater flexibility reduces the amount of 
distortion. \\

\begin{figure}
\includegraphics[12cm,22cm][13cm,23cm]{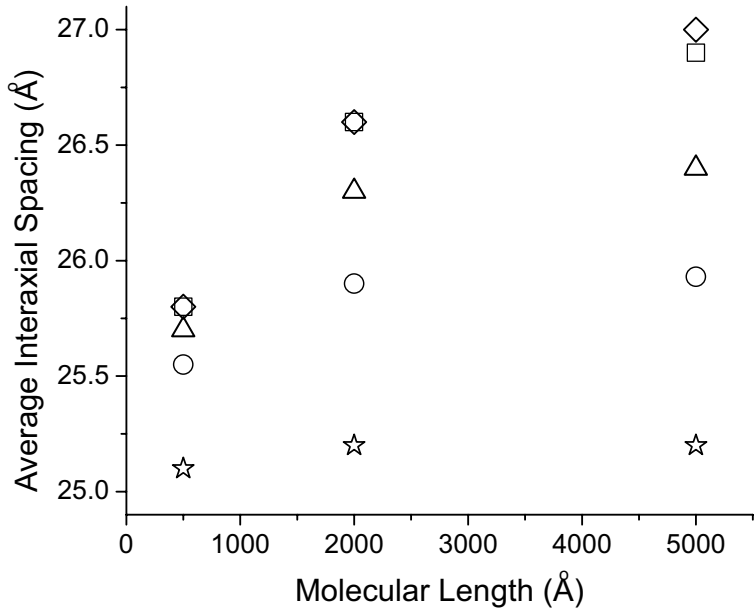}
\vspace{8cm} \caption{
Location of the antiferromagnetic-ferromagnetic transition for molecules 
of different lengths and torsional flexibilities.  At large densities 
(small interaxial spacings), the antiferromagnetic state is the lowest 
energy state whereas the ferromagnetic state is the favored state for more 
dilute assemblies.  Diamonds ($\diamond$) correspond to perfectly 
rigid molecules ($C=\infty$), stars ($\star$) to molecules with a 
torsional modulus of $C=3.0 \times 10^{-19}$ erg-cm, circles ($\circ$) 
to molecules with $C=1.0 \times 10^{-18}$ erg-cm, triangles ($\triangle$) 
to molecules with $C=3.0 \times 10^{-18}$ erg-cm, and squares ($\square$) to 
molecules with $C=3.0 \times 10^{-17}$ erg-cm.  Results are shown for 
molecules with 70\% charge compensation and a 70/30 major/minor groove 
charge distribution in a solution with an inverse Debye screening length 
of about $7 \AA$.
}
\end{figure}

\begin{figure}
\includegraphics[9cm,22cm][10cm,23cm]{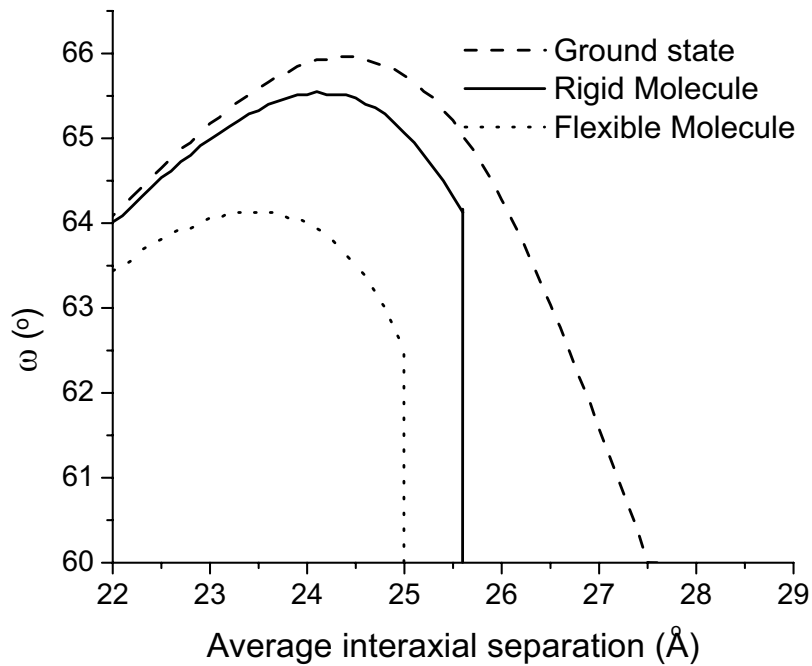}
\vspace{9cm} \caption{The level of distortion ($\omega$) from the 
hexagonal lattice for the antiferromagnetic state as a function of the 
average interaxial separation of the molecules in the columnar 
assemblies.  The distortion is shown for the ground state configuration 
(no spin fluctuations), for a rigid molecule of length 500$\AA$ with 
spin fluctuations, and for a flexible molecule ($C=3.0 \times 10^{-19}$ 
erg-cm) with torsional fluctuations of the same length.
}
\end{figure}

Modifications of the results of the self-consistent Hartree approximation 
when incorporating corrections for independently rotating ends of the 
molecules are quite involved and are left to the appendices.  When including 
these corrections, however, we find that the locations of the transition 
only shift by hundredths of angstroms for all the cases shown in Fig. 3.  
And so the formulation assuming periodic boundary conditions, as long as 
the molecules are long enough, adequately describes the system. \\

We may also calculate correlation functions within the Hartree approximation 
for periodic boundary conditions.  Here, we may use the results of the 
previous section, but we now replace $ \lambda _1 $
with $ \lambda _1^H $ and $\alpha$ with 
$ \alpha _H  = (\lambda _1^H /\lambda _2^H )^2$. We then may compute 
$\Delta_\phi$ and $\Delta_\psi$ within this approximation.  A plot of 
these quantities is shown in Fig. 5 for a molecule with a realistic value 
for its torsional modulus.  The correction associated with fluctuations 
of the free ends of the molecules only results in small changes in the value of
$\Delta _\phi$. (The formulation of which is again left to the appendices.)  
On this plot, we also show the variation of the relative azimuthal 
angle due to rigid body fluctuations, $\Delta _\psi$, for a molecule of the 
same length.  These results are consistent with fits of x-ray diffraction 
data of hydrated DNA assemblies \cite{kornyshev:05}. 
\\

\begin{figure}
\includegraphics[9cm,22cm][10cm,23cm]{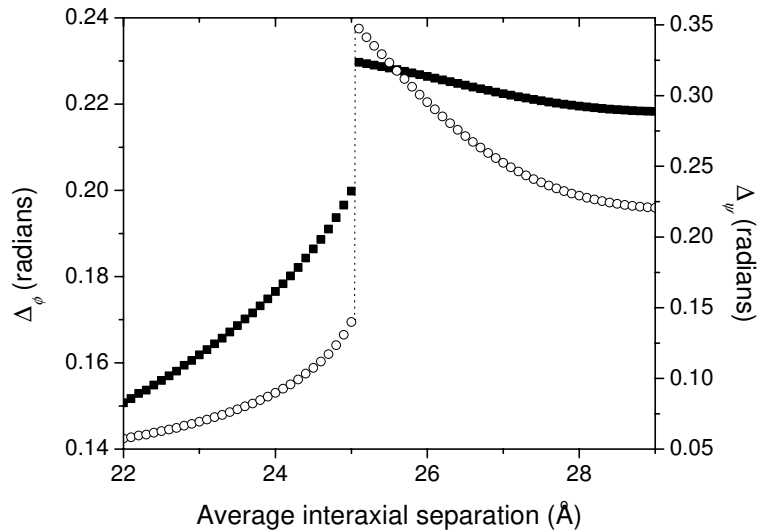}
\vspace{7cm} \caption{The contributions of torsional fluctuations 
$\Delta_{\phi}$ 
(filled sqares) and rigid body fluctuations $\Delta_{\psi}$ 
(open circles) to the 
correlation function of the relative azimuthal angles between molecules 
in an assembly as a function of the average molecular separation.  
Again the break at 25.1$\AA$ corresponds to the transition from the 
antiferromagnetic state (at smaller molecular separations) to the 
ferromagnetic state.  Values are shown for an assembly of molecules 
that are 500$\AA$ long and have torsional moduli of $3.0 \times 10^{-19}$ 
erg-cm.}
\end{figure}

Lastly, we show in Appendix H the high temperature expansion for the 
azimuthally disordered or Kosterlitz-Thouless vortex phase, mentioned in 
the introduction, that includes torsional fluctuations.  Granted, in the 
case of real DNA it is very unlikely that this situation is realized 
\cite{wynveen:05a}.  At larger separations where thermal fluctuations 
dominate intermolecular azimuthal interactions,  the DNA assembly is likely 
to lose columnar ordering before such a transition in the azimuthal 
configuration of the columnar assembly occurs.  Because of this reservation, 
this work has not been included in the main text.  However, in other helical 
bio-molecule assemblies where azimuthal interactions might be weaker but 
spatial interactions are strong, a  Kosterlitz-Thouless like transition 
could possibly occur within a smectic layer with short range 
two-dimensional ordering.
\\

%%%%%%%%%%%%%%%%%%%%%%%%%%%%%%%%%%%%%%%%%%%%%%%%%%%%%%%%%%%%%%%%%%%%%%%%%%

\renewcommand{\theequation}{5.\arabic{equation}}
\setcounter{equation}{0}
\section{Discussion and conclusions}
We have developed a method for calculating the contribution to the free 
energy due to azimuthal fluctuations for torsionally flexible molecules 
of finite length.  Using this method we can determine the influence of 
azimuthal flexibility of molecules within a DNA columnar assembly on the 
position of the phase transition between the antiferromagnetic state 
where the hexagonal lattice is distorted and the ferromagnetic state for 
various molecular lengths.  For molecules with torsional moduli of $
C = 3 \times 10^{ - 19} {\rm erg - cm}$, the approximate value deduced 
from experiments \cite{crothers:92}, torsional flexibility reduces the 
separation at which this occurs compared to that found for the case of 
rigid molecules.  We see that the effects of torsional fluctuations are 
more pronounced for assemblies of long molecules, whereas the results for 
short molecules are little different from those rigid body calculations 
of Ref. \cite{wynveen:05a}, thereby showing that  shortening the molecule 
is nearly equivalent to making the molecule less flexible.   
\\

We found that torsional fluctuations considerably reduce the degree of 
distortion, for long molecules in the antiferromagnetic state, as compared 
to that found for calculations for rigid molecules of the same length.  
Since flexibility allows for readjustment of the azimuthal coordinate 
along the length of the molecules in the lattice, less distortion is required 
to minimize the free energy of the assembly due to the antiferromagnetic 
coupling of the pair potential.  Hence, as compared to ground state 
\cite{harreistot} or finite-temperature rigid body calculations 
\cite{wynveen:05a}, X-ray diffraction patterns must have even better 
resolution to pick out this distortion \cite{strey:00}.
\\

 We were also able to calculate the asymptotic form of the correlation 
function $
\left\langle {\exp \left( {in\left( {\phi _{jl} (z) - 
\phi _{j'l'} (z')} \right)} \right)} \right\rangle 
$ including both rigid body and torsional fluctuations.  This term is found 
in the formulation of the intensity of x-ray diffraction patterns of 
assemblies \cite{kornyshev:05} and is therefore a direct means by 
which one can determine the extent of these fluctuations.  Our results 
for the parameters chosen agree quite well with the fits made to x-ray 
diffraction data of hydrated DNA assemblies.  Conversely, comparing 
these fits to the correlation function formulation may reveal the 
strength of the azimuthal interaction, which in turn may say 
something about the charge distribution on the DNA molecules in the assembly.
\\

Although we have discovered a number of effects associated with allowing 
for  torsional flexibility in columnar DNA assemblies, this analysis, 
however, is limited by the form of the DNA-DNA interaction 
\cite{kornyshev:97a} we have employed.  The effects of nonlocal 
polarizability of the water in the narrow interstitial regions between 
the DNA could alter the results \cite{medvedev:04,kornyshev:99b,solvation}.  
Also, the potential assumes a given charge distribution of readsorbed 
counterions that changes very little with density.  Once these effects 
have been incorporated into the pair potential the same analysis developed 
here may be used.  But even upon considering these effects, the qualitative 
aspects of this analysis should remain unchanged.  
\\

One may also need to consider the effect of sequence dependent distortions 
from the ideal double helical structure seen in real DNA 
\cite{kabsch:82,gorin:95,kornyshev:01a,kornyshev:04a} 
in the analysis of the assemblies.  This may well be included in 
calculations for assemblies of non-homologous DNA by employing ensemble 
averaging  \cite{kornyshev:01a}.  Such averaging is easiest in an 
assembly made up of with several different DNA sequences.  Again, as 
in Ref. \cite{lee:04}, these effects may be incorporated into a combined 
persistence length that incorporates both distortions arising from 
thermal fluctuations and from different base pair sequences of neighboring 
molecules.  For assemblies containing DNA sequences with only a few 
different texts (and thus different distortions from an ideal helix), 
the interaction is much more difficult to treat and has yet been 
considered. Nevertheless, in the case of homologous DNA assemblies, 
distortions from an ideal helix make little difference as the 
electrostatic interaction depends only on the relative azimuthal 
orientations between the molecules.
\\

Finally, we should point out that this is by no means a complete 
statistical mechanical picture of a columnar assembly.  For example, 
we have not included lattice vibrations, fluctuations in the $z$-position of 
the center of each molecule, bending fluctuations, nor fluctuations due 
to other geometrical distortions of each molecule.  However, 
fluctuations in the azimuthal degrees of freedom are likely the most 
important feature for determining the structure of the mesophases of 
columnar assemblies as seen in the fits of x-ray diffraction patterns 
\cite{kornyshev:05}.  Furthermore, coupling of the azimuthal degrees of 
freedom to other geometrical distortions, which are likely to be small, 
may be incorporated into our model. 
\\

\section*{Acknowledgements}
We would like to thank A. A. Kornyshev for helpful discussions.
Financial support from EPSRC grant No. GR/S31068/01 and the Royal Society 
are gratefully acknowledged.
%\end{acknowledgments}

\renewcommand{\theequation}{A\arabic{equation}}
\setcounter{equation}{0}
\section*{Appendix A. Details of the ``Quantum Mechanical'' 
calculation of the free energy at finite length.}
We start with equation. (2.5) of the text, namely
\begin{equation}
F =  - k_B T\ln \sum\limits_{E, + , - } {\left\langle {{\phi _ +  }}
 \mathrel{\left | {\vphantom {{\phi _ +  } E}}
 \right. \kern-\nulldelimiterspace}
 {E} \right\rangle } \exp \left( { - \frac{{EL}}{{\lambda _p }}} \right)\left\langle {E}
 \mathrel{\left | {\vphantom {E {\phi _ -  }}}
 \right. \kern-\nulldelimiterspace}
 {{\phi _ -  }} \right\rangle  - k_B T\ln \Theta. 
\end{equation}

We will now, using Eq. (D1), derive expressions for finite size correction terms for the Gaussian approximation. But first we will need to derive some general results. 
\\
	
Now each Eigen state of Eq. (2.3) is characterised by two numbers; the band number $ b \ge 0 $, an integer, and the wave number in the band $1/2<k<1/2$, a continuous variable. Each Eigen state should then be written as
\begin{equation}
\psi _{b,k} (\phi ) = \left\langle {\phi }
 \mathrel{\left | {\vphantom {\phi  {E_{b,k} }}}
 \right. \kern-\nulldelimiterspace}
 {{E_{b,k} }} \right\rangle  = \mathop {\lim }\limits_{N \to \infty } \frac{1}{{\sqrt {N_P } }}\sum\limits_n {G_{n,b,k} \exp \left( {i(n + k)\phi } \right)} ,
\end{equation}
where the $G_{n.b,k}$ satisfy 
\begin{equation}
\left[ {\frac{{(n + k)^2 }}{2} - E_{b,k} } \right]G_{n,b,k}  - \frac{{\lambda _p^2 }}{{2\lambda _0^2 }}\left[ {G_{n - 1,b,k}  + G_{n + 1.b,k} } \right] + \frac{{a_2 \lambda _p^2 }}{{2a_1 \lambda _0^2 }}\left[ {G_{n - 2,b,k}  + G_{n + 2.b,k} } \right] = 0.
\end{equation}
Here, $N_P$ is the number of periods the potential has in $\phi$-space and is proportional to the system size: the range of values which $\phi$ is allowed to take. As $\phi$ will fluctuate between $- \infty $ and $ \infty $, except at $ \phi _ + $, we take $N_P$ to be infinite. It is easy, then, to show that the following is true
 \begin{equation}
\int\limits_0^{2\pi } {\psi _{b,k} (\phi )d\phi  = \mathop {\lim }\limits_{N_p  \to \infty } } \frac{{2\pi }}{{\sqrt {N_P } }}\delta _{k,0} G_{0,b,0} .
\end{equation}
Using Eq. (D4) we rewrite $F$ as 
\begin{equation}
F =  - k_B T\ln \sum\limits_b {\left| {G_{0,b,0} } \right|^2 } \exp \left( { - \frac{{E_b L}}{{\lambda _p }}} \right) - k_B T\ln \Theta. 
\end{equation}
In the Gaussian approximation we assume that $\psi_{b,0}$ can be written as a linear supposition of simple harmonic oscillator Eigen states
\begin{equation}
\psi _{b,0}  = \mathop {\lim }\limits_{N_P  \to \infty } \frac{{c_b }}{{\sqrt {N_P } }}\sum\limits_{j =  - N_P /2}^{N_P /2} {H_b \left( {\sqrt \omega  (\phi  - 2\pi j)} \right)\exp \left( {\frac{{ - \omega \left( {\phi  - 2\pi j} \right)^2 }}{2}} \right)} ,
\end{equation}
where the $H_b(x)$ are the Hermite polynomials and the $c_b$ are the normalization constants for each state. In the Gaussian approximation  
\begin{eqnarray}
\omega ^2  = \left( {\frac{{\lambda _p^2 }}{{\lambda _0^2 }}} \right)\cos \phi _0  - \left( {\frac{{4a_2 \lambda _p^2 }}{{a_1 \lambda _0^2 }}} \right)\cos 2\phi _0 
\, \, \, {\rm { and}} \nonumber \\
E_b  = \left( {\frac{1}{2} + b} \right)\omega  - \left( {\frac{{\lambda _p^2 }}{{\lambda _0^2 }}} \right)\cos (\phi _0 ) + \left( {\frac{{a_2 \lambda _p^2 }}{{a_1 \lambda _0^2 }}} \right)\cos (2\phi _0 ).
\end{eqnarray}
Assuming negligible overlap between the wave-functions in the superposition; we find on retaining the first two terms in the sum in Eq. (D5) and expanding out the logarithm and writing $\omega=\lambda_p/\lambda$ we find
\begin{equation}
F = E_0 + \frac{{CL}}{{4\lambda _p \lambda }} - \frac{{k_B T}}{2}\exp 
\left( { - \frac{{(E_2  - E_0 )L}}{{\lambda _p }}} \right) + \frac{{k_B T}}{2}\ln \left( {\frac{\lambda }{{\lambda _p }}} \right) - k_B T\ln \Theta. 
\end{equation}
We should point out that this result differs from (D9) of \cite{lee:04}, 
as the previous version a couple of mistakes that have been corrected. 
Substituting for $E_0$ and $E_2$ we so obtain Eq. (2.6) of the text.

\renewcommand{\theequation}{B\arabic{equation}}
\setcounter{equation}{0}
\section*{Appendix B. Calculation of the free energy for the DNA pair by Gaussian integration.}
 We may readily perform the Gaussian integrations in Eq.(2.10) of the text 
\begin{equation}
Z = Z_0 Z_f Z_p \exp \left( { - \frac{{E_0 }}{{k_b T}}} \right),
\end{equation}
where now
\begin{eqnarray}
Z_0  = \left( {\frac{{2\pi k_B T}}{m}} \right)^{1/2}, \nonumber \\ 
Z_f  = \left( {\frac{{4\pi \lambda }}{{\lambda _p }}\tanh \left( {\frac{L}{{2\lambda }}} \right)} \right)^{1/2}, \nonumber \\
Z_p  = \prod\limits_{n \ne 0} {\left( {2\pi k_B TG{}_n} \right)^{1/2} }. 
\end{eqnarray}
Using  the definition of free energy $ F =  - k_B T\ln Z $ we are then able to write the free energy as
 \begin{equation}
F = \frac{{k_B T}}{2}\sum\limits_n {\ln \left( {\frac{C}{2}\left( {\frac{{2\pi n}}{L}} \right)^2  + m} \right) + \frac{{k_B T}}{2}\ln \left( {\frac{{\lambda _p }}{{2\lambda }}\coth \left( {\frac{L}{{2\lambda }}} \right)} \right)}  + k_B T\Theta, 
\end{equation}
where we have adsorbed terms that dont depend on $\lambda$ and are not important to our
analysis into $\Theta$. The first term may be evaluated by means of a trick
\begin{equation}
\frac{\partial }{{\partial m}}\sum\limits_n {\ln \left( {\frac{C}{2}\left( {\frac{{2\pi n}}{L}} \right)^2  + m} \right)}  = \sum\limits_n {\left( {\frac{C}{2}\left( {\frac{{2\pi n}}{L}} \right)^2  + m} \right)^{ - 1} }  = \frac{{L\lambda }}{C}\coth \left( {\frac{L}{{2\lambda }}} \right).
\end{equation}
We then by integrate up (B3) and so arrive at 
\begin{equation}
F = k_B T\ln \left( {\sinh \left( {\frac{L}{{2\lambda }}} \right)} \right) + \frac{{k_B T}}{2}\ln \left( {\frac{{\lambda _p }}{{2\lambda }}\coth \left( {\frac{L}{{2\lambda }}} \right)} \right) + k_B T\Theta, 
\end{equation}
where extra constants, such as constants of integration have been adsorbed into $\Theta$.
On rearrangement we arrive at (2.11) in the text.

%%%%%%%%%%%%%%%%%%%%%%%%%%%%%%%%%%%%%%%%%%%%%%%%%%%%%%%%%%%%%%%%%%%%

\renewcommand{\theequation}{C\arabic{equation}}
\setcounter{equation}{0}
\section*{Appendix C. Calculation of the  Free Energy for Gaussian fluctuations in the assembly.}
We start our analysis by substituting (3.7) into (3.5) of the text and so obtain the following expression for $E_H^{AF} [\phi ']$ 
\begin{eqnarray}
E_H^{AF} [\phi '] &=& \sum\limits_{jl} \int_{ - L/2}^{L/2} {dz} 
\left[ 
\frac{C}{2}\left( {\frac{{d\phi _{jl}^p }}{{dz}}} \right)^2  + 
\frac{{C\gamma _{jl}^2 }}{{2L^2 }} \right. \nonumber \\
&+& \frac{{m_1 }}{2} \left[ 
{\left( {\phi _{jl}^p  - \phi _{j - 1l}^p  + \frac{{z\gamma _{jl} }}{L} - 
\frac{{z\gamma _{j - 1l} }}{L}} \right)^2  + \left( {\phi _{jl}^p  - 
\phi _{jl - 1}^p  + \frac{{z\gamma _{jl} }}{L} - 
\frac{{z\gamma _{jl - 1} }}{L}} \right)^2 } \right] \\
 &+& \frac{{m_2 }}{2}\left. {\left( {\phi _{jl}^p  - 
\phi _{j + 1l - 1}^p  + \frac{{z\gamma _{jl} }}{L} - 
\frac{{z\gamma _{j + 1l - 1} }}{L}} \right)^2 } \right]
. \nonumber
\end{eqnarray}
We then introduce the following lattice Fourier transforms
\begin{equation}
\phi _{jl}^p (z) = \frac{1}{{\sqrt A }}\sum\limits_{k_u k_v } {\phi ^p (\vec k,z)e^{i(jk_u  + lk_v )r_0 } } \,\, {\rm      and}\,\,
\gamma _{jl}  = \frac{1}{{\sqrt A }}\sum\limits_{k_u k_v } {\gamma (\vec k)e^{i(jk_u  + lk_v )r_0 } }, 
\end{equation}
where $A$ is total area of the lattice and $r_0$ is the lattice spacing.  
We have chosen reciprocal lattice vectors corresponding to the rhombic 
Bravais lattice defined by $ \hat u$ and $ \hat v$ see Fig. 1 of the text. Then $k_u$ and $k_v$ take on values which lie within the first Brillouin zone for a rhombic lattice (e.g. $-\pi/r_0 < k_u < \pi/r_0$). We are then able to write in the limit $ A \to \infty $
\begin{eqnarray}
E_H^{AF} [\phi ] = \frac{1}{{(2\pi )^2 }}\int\limits_{ - \pi }^\pi  {dx} \int\limits_{ - \pi }^\pi  {dy} \int_{ - L/2}^{L/2} {dz} \left[ {\frac{C}{2}\left( {\frac{{\partial \phi ^p (x,y,z)}}{{\partial z}}\frac{{\partial \phi ^p ( - x, - y,z)}}{{\partial z}}} \right) + \frac{{C\gamma (x,y)\gamma ( - x, - y)}}{{2L^2 }}} \right. \nonumber \\
 + m_1 \hat C(x,y;\alpha )\left. {\left[ {\left( {\phi ^p (x,y,z) + \frac{{z\gamma (x,y)}}{L}} \right)\left( {\phi ^p ( - x, - y,z) + \frac{{z\gamma ( - x, - y)}}{L}} \right)} \right]} \right],
\end{eqnarray}
where $x=r_0k_u$ and $y=r_0k_v$. We then may express $\phi^p(x,y,z)$ through Fourier Series
\begin{equation}
\phi ^p (x,y,z) = \frac{1}{{\sqrt L }}\sum\limits_{n = 0} {b_n (x,y)\exp \left( {\frac{{2\pi inz}}{L}} \right)}, 
\end{equation}
and so re-express (C3)
\begin{eqnarray}
E_H^{AF} [\phi ] & = & 
\frac{1}{{(2\pi )^2 }}\int\limits_{ - \pi }^\pi  {dx} 
\int\limits_{ - \pi }^\pi  dy\left( \sum\limits_n \left[ 
{\frac{{b_n (x,y)b_{ - n} ( - x, - y)}}{{2G_n (x,y)}}} \right. \right. 
\nonumber \\
&+& 
\left. m\hat C(x,y;\alpha )\gamma (x,y)J_n b_{ - n} ( - x, - y) + 
m\hat C(x,y;\alpha )\gamma ( - x, - y)J_{ - n} b_n (x,y) 
\right] \\
&+& \left. \left( {\frac{C}{{L^2 }}} + 
\frac{{Lm}}{6}\hat C(x,y;\alpha  \right)\frac{{\gamma (x,y)\gamma 
( - x, - y)}}{{2L^2 }} \right), \nonumber
\end{eqnarray}
where 
$
G_n (x,y) = \left( {C\left( {\frac{{2\pi n}}{L}} \right)^2  + 2m\hat C(x,y;\alpha )} \right)^{ - 1} 
$, $ J_n  = \frac{{i( - 1)^n \sqrt L }}{{2\pi n}}\left( {1 - \delta _{n,0} } \right) $, $
\hat C(x,y;\alpha ) = (1 - \cos (x)) + (1 - \cos (y)) + \alpha (1 - \cos (x - y)) $ and $ \alpha=m_2/m_1 $. On making the variable shift $ b_n (x,y) \to b_n (x,y) - m_1 \gamma (x,y)\hat C(x,y;\alpha )J_n G_n (x,y) $ we may write this as
\begin{eqnarray}
E_H^{AF} [\phi ] = \frac{1}{{(2\pi )^2 }}\int\limits_{ - \pi }^\pi  {dx}  
\int\limits_{ - \pi }^\pi  
& dy & \left(  \sum\limits_n {\left[ 
{\frac{{b_n (x,y)b_{ - n} ( - x, - y)}}{{2G_n (x,y)}} 
+ 2m^2 \hat C(x,y;\alpha )^2 
\gamma (x,y)\gamma ( - x, - y)J_{ - n} G_n (x,y)J_n } 
\right]}  \right. \nonumber \\
 &+& \left. \left( {\frac{C}{L} + \frac{{Lm}}{6}\hat C(x,y;\alpha )} 
\right)\frac{{\gamma (x,y)\gamma ( - x, - y)}}{{2L^2 }} \right).
\end{eqnarray}
We find that on further manipulation we may write
\begin{equation}
Z = Z_f Z_p \exp \left( { - \frac{{E_0^{AF} }}{{k_B T}}} \right),
\end{equation}
where
\begin{eqnarray}
Z_f  = \int {{\cal D}\gamma } (x,y)\exp \left( { - \frac{1}{{(2\pi )^2 }}\int\limits_{ - \pi }^\pi  {dx\int\limits_{ - \pi }^\pi  {dy} \frac{{\gamma (x,y)\gamma ( - x, - y)}}{{2S(x,y)}}} } \right), \nonumber \\
Z_p  = \prod\limits_n {\int {\cal D} b_n (x,y)} \exp \left( { - \frac{1}{{(2\pi )^2 }}\sum\limits_n {\int\limits_{ - \pi }^\pi  {dx} } \int\limits_{ - \pi }^\pi  {dy\frac{{b_n (x,y)b_{ - n} ( - x, - y)}}{{2G_n (x,y)k_B T}}} } \right), \nonumber
\end{eqnarray}
and $ S(x,y) = \frac{\lambda }{{\lambda _p }}\frac{1}{{\hat C(x,y;\alpha )^{1/2} }}\tanh \left( {\frac{L}{{2\lambda }}\hat C(x,y;\alpha )^{1/2} } \right)$. We may evaluate the path integrals
giving
\begin{eqnarray}
Z_f  = \mathop {\lim }\limits_{\mathop {N_x  \to \infty }\limits_{N_y  \to \infty } } \prod\limits_{n_x ,n_y } {\left( {2\pi S\left( {\frac{{n_x }}{{2\pi N_x }},\frac{{n{}_y}}{{2\pi N_y }}} \right)} \right)^{1/2} } 
\nonumber \\
Z_p  = \prod\limits_n {\mathop {\lim }\limits_{\mathop {N_x  \to \infty }\limits_{N_y  \to \infty } } \prod\limits_{n_x ,n_y } {\left( {2\pi k_B TG_n \left( {\frac{{n_x }}{{2\pi N_x }},\frac{{n{}_y}}{{2\pi N_y }}} \right)} \right)^{1/2} } }. 
\end{eqnarray}
The product $N=N_x N_y$ is the number of molecules in the assembly which we have taken to be infinite as 
$ A \to \infty $. From (C8) we find the following result for the free energy per molecule 
\begin{eqnarray}
F =  - \frac{{k_B T}}{{2(2\pi )^2 }}\int\limits_{ - \pi }^\pi  {dx} \int\limits_{ - \pi }^\pi  {dy\ln \left( {S(x,y)} \right)}  + \frac{{k_B T}}{{2(2\pi )^2 }}\sum\limits_n {\int\limits_{ - \pi }^\pi  {dx} } \int\limits_{ - \pi }^\pi  {dy} \ln \left( {C\left( {\frac{{2\pi n}}{L}} \right)^2  + 2m_1 \hat C(x,y;\alpha )} \right) \nonumber \\
 + k_B T\Theta _{asb}, 
\end{eqnarray}
where unimportant terms that do not depend on $m_1$ or $\alpha$ have been adsorbed into $\Theta_{asb}$. 
Using the trick illustrated by Eq. (B3) It is possible to write
\begin{eqnarray}
F = \frac{{k_B T}}{{2(2\pi )^2 }}\int\limits_{ - \pi }^\pi  {dx} \int\limits_{ - \pi }^\pi  {dy\ln \left( {\frac{{\lambda _p }}{\lambda }\left( {\frac{{\hat C(x,y;\alpha )}}{{\hat C(x,y;1)}}} \right)^{1/2} \coth \left( {\frac{{\hat C(x,y;\alpha )^{1/2} L}}{{2\lambda }}} \right)} \right)} 
\nonumber \\
 + \frac{{k_B T}}{{(2\pi )^2 }}\int\limits_{ - \pi }^\pi  {dx} \int\limits_{ - \pi }^\pi  {dy} \ln \left( {\frac{1}{{\hat C(x,y,1)}}\sinh \left( {\frac{L}{{2\lambda }}\hat C(x,y,\alpha )} \right)} \right) + k_B T\Theta _{asb}, 
\end{eqnarray}
where additional constants have been adsorbed into $\Theta_{asb}$.

\renewcommand{\theequation}{D\arabic{equation}}
\setcounter{equation}{0}
\section*{Appendix D. Correlation functions for Gaussian fluctuations in the assembly.}
It is first useful to consider the following correlation function
\begin{equation}
G(j - j',l - l',z,z') = \left\langle {\phi '_{jl} (z)\phi '_{j'l'} (z')} \right\rangle _0  - \frac{1}{2}\left\langle {\phi '_{jl} (z)^2 } \right\rangle _0  - \frac{1}{2}\left\langle {\phi '_{j'l'} (z')^2 } \right\rangle _0, 
\end{equation}
where in general for any quantity $A[\phi ]$
\begin{equation}
\left\langle {A[\phi ]} \right\rangle _0  = \exp \left( { - \frac{{E_0^{AF} }}{{k_B T}}} \right){\rm  }\frac{1}{Z}\prod\limits_{jl} {\int {\cal D} \phi _{jl} (z)A[\phi ]} \exp \left( { - \frac{{E_H^{AF} [\phi ]}}{{k_B T}}} \right),
\end{equation}
using the ansatz (3.7) we may write (D1) as
\begin{eqnarray}
G(j - j',l - l',z,z') = \left\langle {\phi _{jl}^p (z)\phi _{j'l'}^p (z')} \right\rangle _0  - \frac{1}{2}\left\langle {\phi _{jl}^p (z)^2 } \right\rangle _0  - \frac{1}{2}\left\langle {\phi _{j'l'}^p (z')^2 } \right\rangle _0 
\nonumber \\
 + \frac{z}{L}\left[ {\left\langle {\gamma _{jl} \phi _{j'l'}^p (z')} \right\rangle _0  - \left\langle {\gamma _{jl} \phi _{jl}^p (z)} \right\rangle _0 } \right] + \frac{{z'}}{L}\left[ {\left\langle {\phi _{jl}^p (z)\gamma _{j'l'} } \right\rangle _0  - \left\langle {\phi _{j'l'}^p (z')\gamma _{j'l'} } \right\rangle _0 } \right]
\\
 - \frac{{z^2 }}{{2L^2 }}\left\langle {\gamma _{jl}^2 } \right\rangle _0  - \frac{{z'^2 }}{{2L^2 }}\left\langle {\gamma _{j'l'}^2 } \right\rangle _0  + \frac{{zz'}}{{L^2 }}\left\langle {\gamma _{jl} \gamma _{j'l'} } \right\rangle _0. \nonumber
\end{eqnarray}
This may be rewritten as
\begin{eqnarray}
G(j - j',l - l',z,z') &=& \frac{1}{{\left( {2\pi } 
\right)^2 L}}\sum\limits_{n,n'} {\int\limits_{ - \pi }^\pi  {dx} 
\int\limits_{ - \pi }^\pi  {dy} } \left( \left[ {\left\langle {b_n (x,y)b_{n'} ( - x, - y)} \right\rangle _0 } \right. \right.
\\
&+& J_n \left\langle {\gamma (x,y)b_{n'} ( - x, - y)} \right\rangle _0  + J_{n'} \left\langle {\gamma ( - x, - y)b_n (x,y)} \right\rangle _0 \left. { + J_n J_{n'} \left\langle {\gamma ( - x, - y)\gamma (x,y)} \right\rangle _0 } \right]
\nonumber \\
& \times &  
\left( \exp \left( {i(j - j')x + i(l - l')y + \frac{{2i\pi (nz + n'z')}}{L}} 
\right) - \frac{1}{2}\exp \left( {\frac{{2i\pi (n + n')z}}{L}} \right) \right. 
\nonumber \\
&-& \left. \left. \frac{1}{2}\exp \left( 
{\frac{{2i\pi (n + n')z'}}{L}} \right) \right) \right) .
\nonumber
\end{eqnarray}
We then make the variable shift $ b_n (x,y) \to b_n (x,y) - m_1 \gamma (x,y)\hat C(x,y;\alpha )J_n G_n (x,y) $ and obtain the following 
\begin{equation}
G(j - j',l - l',z,z') = \tilde G_P (j - j',l - l',z - z') + G_{AP} (j - j',l - l',z,z'),
\end{equation}
where
\begin{equation}
\tilde G_P (j,l,z) = \frac{{k_B T}}{{(2\pi )^2 L}}\sum\limits_n {\int\limits_{ - \pi }^\pi  {dx} } \int\limits_{ - \pi }^\pi  {dy} G_n (x,y)\left( {\exp \left( {i(jx + ly) + \frac{{2i\pi nz}}{L}} \right) - 1} \right),
\end{equation}
and
\begin{eqnarray}
G_{AP} (j,l,z,z') &=& \frac{1}{{(2\pi )^2 L}}\sum\limits_{n,n'} {\int\limits_{ - \pi }^\pi  {dx} } \int\limits_{ - \pi }^\pi  {dy} \left( S(x,y) \right.
 \\
& \times & C^2 \left( {\frac{{(2\pi )^2 nn'}}{{L^2 }}} \right)( - 1)^{n + n'} \left( {C\left( {\frac{{2\pi n}}{L}} \right)^2  + 2m\hat C(x,y;\alpha )} \right)^{ - 1} \left( {C\left( {\frac{{2\pi n'}}{L}} \right)^2  + 2m\hat C(x,y;\alpha )} \right)^{ - 1} 
\nonumber \\
& \times & \left.
\left( {\exp \left( {i(jx + ly) + \frac{{2i\pi (nz + n'z')}}{L}} \right) - \frac{1}{2}\exp \left( {\frac{{2i\pi (n + n')z'}}{L}} \right) - \frac{1}{2}\exp \left( {\frac{{2i\pi (n + n')z}}{L}} \right)} \right) \right. .
\nonumber
\end{eqnarray}
In these expressions we are able to perform the sum and we also separate out the rigid body contribution by writing $\tilde G_P (j,l,z) = G_P (j,l,z) + G_0 (j,l)$ where $G_0 (j,l)$ is the part of $G(j,l,z,z')$
that arises purely from rigid body fluctuations.  We obtain the following results
\begin{equation}
G_0 (j,l) = \frac{{k_B T}}{{(2\pi )^2 2Lm_1 }}\int\limits_{ - \pi }^\pi  {dx} \int\limits_{ - \pi }^\pi  {dy} \frac{1}{{\hat C(x,y;\alpha )}}\left( {\exp (i(jx + ly)) - 1} \right),
\end{equation}
\begin{eqnarray}
G_P (j,l,z) = \frac{{k_B T}}{{2C}}\frac{1}{{(2\pi )^2 }}\int\limits_{ - \pi }^\pi  {dx} \int\limits_{ - \pi }^\pi  {dy} \left[ {\left\{ {\frac{{\lambda _1 }}{{\hat C(x,y;\alpha )^{1/2} }}\left( {\exp \left( { - \frac{{\left| z \right|\hat C(x,y;\alpha )^{1/2} }}{{\lambda _1 }}} \right)} \right.} \right.} \right.
\nonumber \\
\left. { + \exp \left( {\frac{{(\left| z \right| 
- L)\hat C(x,y;\alpha )^{1/2} }}{\lambda _1 }} \right)} \right)\left. 
{\left( {1 - \exp \left( 
{\frac{{ - L\hat C(x,y;\alpha )^{1/2} }}{\lambda _1 }} 
\right)} \right)^{ - 1}  - \frac{{2\lambda _1^2 }}{{L\hat C(x,y;\alpha )}}} \right\}
\nonumber \\
\left. {\exp \left( {i(jx + ly)} \right) - \left( {\frac{{\lambda _1 }}{{\hat C(x,y;\alpha )^{1/2} }}\coth \left( {\frac{{L\hat C(x,y;\alpha )^{1/2} }}{{2\lambda _1 }}} \right) - \frac{{2\lambda _1^2 }}{{L\hat C(x,y;\alpha )}}} \right)} \right],
\nonumber
\end{eqnarray}
\begin{eqnarray}
G_{AP} (j,l,z,z') = \frac{{k_B T\lambda _1 }}{2C}\frac{1}{{(2\pi )^2 }}\int\limits_{ - \pi }^\pi  {dx} \int\limits_{ - \pi }^\pi  {dy} \frac{1}{{\hat C(x,y;\alpha )^{1/2} }}\exp \left( { - \frac{{L\hat C(x,y;\alpha )^{1/2} }}{{\lambda _1 }}} \right)\left( {1 - \exp \left( { - \frac{{2L\hat C(x,y;\alpha )^{1/2} }}{{\lambda _1 }}} \right)} \right)^{ - 1} 
\nonumber \\
{{\mathop{\rm sgn}} (z) {\mathop{\rm sgn}} (z')\left( {\exp \left( 
{\frac{{\left| z \right|\hat C(x,y;\alpha )^{1/2} }}{\lambda _1 }} 
\right) - \exp \left( { - \frac{{\left| z 
\right|\hat C(x,y;\alpha )^{1/2} }}{\lambda _1 }} \right)} \right)}
\nonumber \\
\left( {\exp \left( {\frac{{\left| {z'} 
\right|\hat C(x,y;\alpha )^{1/2} }}{\lambda _1 }} \right) - \exp 
\left( { - \frac{{\left| {z'} 
\right|\hat C(x,y;\alpha )^{1/2} }}{\lambda _1 }} \right)} 
\right)\exp \left( {i(jx + ly)} \right)
\nonumber \\
 - \frac{1}{2} \left[ {\left( {\exp \left( 
{\frac{{\left| z \right|\hat C(x,y;\alpha )^{1/2} }}{\lambda _1 }} 
\right) - \exp \left( {\frac{{ - \left| z 
\right|\hat C(x,y;\alpha )^{1/2} }}{\lambda _1 }} \right)} \right)^2 } \right.
\nonumber \\
\left. {\left. { + \left( {\exp \left( {\frac{{\left| {z'} 
\right|\hat C(x,y;\alpha )^{1/2} }}{\lambda _1 }} \right) - 
\exp \left( {\frac{{ - \left| {z'} 
\right|\hat C(x,y;\alpha )^{1/2} }}{\lambda _1 }} \right)} \right)^2 } \right]
} \right].
\nonumber
\end{eqnarray}
Now from these expressions let us find their behavior when $l,j \gg 1$. 
For both $G_{AP}(j,l,z,z')$ and 
$G_{P}(j,l,z)$ this is simple. As $j$ and $l$ increase, the terms 
in the integrands of $G_{AP}(j,l,z,z')$ and $G_{P}(j,l,z)$ that depend 
on $j$ and $l$ oscillate more rapidly, so that their contribution gets 
smaller and smaller. So when $l,j \gg 1$ we may neglect terms that 
depend on $j$ and $l$ in both integrands, so arriving at (3.13) and (3.14)  
of the text. The term $G_0(j,l)$ is trickier as neglecting terms that 
depend on $j$ and $l$   leads to a logarithmic divergence. It is, 
however, possible to do one of the integrations in (D8) 
(by contour integration around a unit circle) which leads to the following result (for $j>0$)
\begin{eqnarray}
G_0 (j,l) = \frac{{k_B T}}{{2(2\pi )Lm_1 }}\int\limits_{ - \pi }^\pi  {dy} \left( {\exp \left( {ily + j\ln \left( {z^ -  (y)} \right)} \right) + \left. {\exp \left( { - ily - j\ln \left( {z^ +  (y)} \right)} \right) - 2} \right)} \right.
\nonumber \\
\frac{1}{{(z^ -  (y) - z^ +  (y))}}\frac{1}{{(1 + \alpha \exp ( - iy))}},
\end{eqnarray}
where 
\begin{equation}
z^ \pm  (y) = \frac{{(2 + \alpha  - \cos (y)) \pm 2\left| {\sin (y/2)} \right|\sqrt {2 + 2\alpha  - \cos ^2 (y/2)} }}{{(1 + \alpha \exp ( - iy))}}.
\end{equation}
Now when $l,j \gg 1$ the integral is dominated by small values of $y$. We may expand out the integrand for small $y$, and so obtain 
\begin{eqnarray}
I(j,l) = G_0 (j,l) - G_0 (1,0) \simeq \frac{{2k_B T}}{{(2\pi )\sqrt {1 + 2\alpha } Lm_1 }}\int\limits_0^\infty  {dy} \left( {\cos \left( {\frac{{\alpha y}}{{1 + \alpha }}} \right)\exp \left( { - \frac{{y\sqrt {1 + 2\alpha } }}{{1 + \alpha }}} \right)} \right. \nonumber \\
\left. { - \cos \left( {ry} \right)\exp \left( { - \frac{{yj\sqrt {1 + 2\alpha } }}{{1 + \alpha }}} \right)} \right)\frac{1}{y},
\end{eqnarray}
where $r = l + \alpha j/(\alpha  + 1)$. 
\\

Here, it is convenient to subtract off $G_0(1,0)$ as we shall see. We may evaluate (D11)  by differentiating (D11), which enables us to perform the resulting integrals, and so we obtain the following differential equations, 
\begin{eqnarray}
\left( {\frac{{\partial I}}{{\partial j}}} \right)_r  = \frac{{2k_B T}}{{(2\pi )Lm_1 (1 + \alpha )}}\frac{{\frac{{\sqrt {1 + 2\alpha } }}{{(1 + \alpha )}}j}}{{\frac{{1 + 2\alpha }}{{(1 + \alpha )^2 }}j^2  + r^2 }},
\nonumber \\
\frac{{\partial I}}{{\partial r}} = \frac{{2k_B T}}{{(2\pi )Lm_1 \sqrt {1 + 2\alpha } }}\frac{r}{{\frac{{1 + 2\alpha }}{{(1 + \alpha )^2 }}j^2  + r^2 }},
\end{eqnarray}
with the boundary condition that $I(1,0)=0$. By solving (D12), subject to the boundary condition, substituting for $r$, we find the following asymptotic form 
\begin{equation}
I = \frac{{k_B T}}{{(2\pi )Lm_1 \sqrt {1 + 2\alpha } }}\ln \left( {j^2  + l^2  + \frac{{2\alpha jl}}{{(1 + \alpha )}}} \right),
\end{equation}
and so we are able to obtain (3.12) of the text. 
\\

We find that able to show that on applying (3.5), (D2) and (C6)  and making the shift $
b_n (x,y) \to b_n (x,y) - m_1 \gamma (x,y)\hat C(x,y;\alpha )J_n G_n (x,y) $ that we may write
\begin{eqnarray}
\left\langle {\exp \left( {in\left( {\phi '_{jl} (z) - \phi '_{j'l'} (z')} \right)} \right)} \right\rangle _0  = \frac{1}{Z}\exp \left( { - \frac{{E_0^{AF} }}{{k_B T}}} \right)\prod\limits_n {\int {\cal D} b_n (x,y)} \int {{\cal D}\gamma (x,y)} 
\nonumber \\
\exp \left( {\sum\limits_n {\int\limits_{ - \pi }^\pi  {dx\int\limits_{ - \pi }^\pi  {dy} \frac{{in}}{{\sqrt L (2\pi )^2 }}\left( {b_n (x,y) + \gamma (x,y)\hat J_n (x,y)} \right)} } } \right)
\\
\exp \left( { - \frac{1}{{(2\pi )^2 }}\int\limits_{ - \pi }^\pi  {dx\int\limits_{ - \pi }^\pi  {dy} \frac{{\gamma (x,y)\gamma ( - x, - y)}}{{2S(x,y)}} - \frac{1}{{(2\pi )^2 }}\sum\limits_n {\int\limits_{ - \pi }^\pi  {dx} } \int\limits_{ - \pi }^\pi  {dy\frac{{b_n (x,y)b_{ - n} ( - x, - y)}}{{2G_n (x,y)k_B T}}} } } \right),
\nonumber 
\end{eqnarray}
where $ \hat J_n (x,y) = J_n (1 - 2m\hat C(x,y;\alpha )G_n (x,y)) $.  Now on completing the square and using (C7) we may show (3.9).

\renewcommand{\theequation}{E\arabic{equation}}
\setcounter{equation}{0}
\section*{Appendix E. Perturbation Theory.}
The interaction energy may be expanded out to beyond quadratic order in $\phi'$ according the prescription given by equations (3.6) and (4.2). 
\begin{equation}
Z = \exp \left( { - \frac{{E_0^{AF} }}{{k_B T}}} \right)\prod\limits_{jl} {\int {\cal D} \phi '(z)} {\rm  }\exp \left( { - \frac{{E_H^{AF} [\phi '] + E_L^{AF} [\phi '] + E_{AH}^{AF} [\phi ']}}{{k_B T}}} \right).
\end{equation}
The first two terms, $E_0^{AF}$ and $E_H^{AF} [\phi ]$ are given in the text. 
The next term, $E_L^{AF} [\phi ]$ comes from linear terms, which  no-longer vanish as (3.2) is no-longer satisfied. This term takes the form
\begin{equation}
E_L^{AF} [\phi '] = \sum\limits_{jl} {\int\limits_{ - L/2}^{L/2} {dz} } \left( {a_1 \sin \left( \psi  \right) - 2a_2 \sin \left( {2\psi } \right)} \right)\left( {\left( {\phi '_{jl} (z) - \phi '_{j - 1l} (z)} \right) + \left( {\phi '_{jl} (z) - \phi '_{jl - 1} (z)} \right)} \right).
\end{equation}
The last term may be split into two pieces $E_{AH}^{AF} [\phi '] = E_{AH}^{(1)} [\phi '] + E_{AH}^{(2)} [\phi ']$ where we have
\begin{eqnarray}
E_{AH}^{(1)} [\phi '] = L\sum\limits_{jl} {\sum\limits_{n = 2}^\infty  {\int\limits_{ - L/2}^{L/2} {dz} \frac{1}{{(2n)!}}\, \times } } \left\{ {\left( {\tilde a_1 ( - 1)^{n - 1}  + \tilde a_2 ( - 4)^n } \right)\left( {\phi '_{j,l} (z) - \phi '_{j + 1,l - 1} (z)} \right)^{2n}  + } \right.
\\
\left. { + \left( {a_1 ( - 1)^{n - 1} \cos (\psi ) + \cos (2\psi )a_2 ( - 4)^n } \right)\left( {\left( {(\phi '_{j,l} (z) - \phi '_{j - 1,l} (z))^{2n}  + (\phi '_{j,l} (z) - \phi '_{j,l - 1} (z))^{2n} } \right)} \right)} \right\},
\nonumber
\end{eqnarray}
\begin{eqnarray}
E_{AH}^{(2)} [\phi '] = L\,\sum\limits_{jl} {\sum\limits_{n = 2}^\infty  {\int\limits_{ - L/2}^{L/2} {dz} \frac{1}{{(2n - 1)!}}} } \left( {a_1 \sin \psi ( - 1)^{n - 1}  + a_2 \sin 2\psi ( - 4)^n /2} \right)
\nonumber \\
\left( {(\phi '_{j,l} (z) - \phi '_{j - 1,l} (z))^{2n - 1}  + (\phi '_{j,l} (z) - \phi '_{j - 1,l + 1} (z))^{2n - 1} } \right).
\nonumber
\end{eqnarray}

Here, and throughout the appendices, we adopt the convention $a_n  = a_n (R_1 )$ and $\tilde a_n  = a_n (R_2 )$. For the moment let us truncate these series at $n=2$ terms. We then use the ansatz  (3.7) and perform the lattice Fourier transforms given by (C2) as well (C4).We then may write
\begin{equation}
E_L^{AF} [\phi '] = \left( {a_1 \sin \left( \psi  \right) - 2a_2 \sin \left( {2\psi } \right)} \right)\sqrt L \mathop {\mathop {\lim }\limits_{\scriptstyle x \to 0 \hfill \atop 
  \scriptstyle y \to 0 \hfill} \left[ {\left( {2 - \exp (ix) - \exp (iy)} \right)b_0 (x,y)} \right]}\limits_{}, 
\end{equation}
\begin{eqnarray}
E_{AH}^{(1)} [\phi '] =  - \frac{1}{{4!L(2\pi )^6 }}\sum\limits_{n,n',n''} {\int\limits_{ - \pi }^\pi  {dx} } \int\limits_{ - \pi }^\pi  {dx'} \int\limits_\pi ^\pi  {dx''} \int\limits_{ - \pi }^\pi  {dy} \int\limits_{ - \pi }^\pi  {dy'} \int\limits_\pi ^\pi  {dy''\left( {g_1 F_1^{(4)} (x,x',x'',y,y',y'')} \right.} 
\nonumber \\
\left. { + g_2 F_2^{(4)} (x,x',x'',y,y',y'')} \right)\left\{ {b_n (x,y)b_{n'} (x',y')} \right.b_{n''} (x'',y'')b_{ - n - n' - n''} ( - x - x' - x'', - y - y' - y'')
\nonumber \\
 + 4b_n (x,y)b_{n'} (x',y')b_{n''} (x'',y'')\gamma ( - x - x' - x'', - y - y' - y'')\hat J_{ - n - n' - n''} ( - x - x' - x'', - y - y' - y'')
\nonumber \\
 + 6b_n (x,y)b_{n'} (x',y')\gamma (x'',y'')\gamma ( - x - x' - x'', - y - y' - y'')\hat J_{n''} (x'',y'')
\nonumber \\
\hat J_{ - n - n' - n''} ( - x - x' - x'', - y - y' - y'') + 4b_n (x,y)\gamma (x',y')\gamma (x'',y'')\gamma ( - x - x' - x'', - y - y' - y'')
\nonumber \\
\hat J_{n'} (x',y')\hat J_{n''} (x'',y'')\hat J_{ - n - n' - n''} ( - x - x' - x'', - y - y' - y'') + \gamma (x,y)\gamma (x',y')\gamma (x'',y'')
\nonumber \\
\gamma ( - x - x' - x'', - y - y' - y'')\hat J_n (x,y)\hat J_{n'} (x',y')\hat J_{n''} (x'',y'')\hat J_{ - n - n' - n''} ( - x - x' - x'', - y - y' - y''),
\nonumber
\end{eqnarray}
\begin{eqnarray}
E_{AH}^{(2)} [\phi '] =  - \frac{1}{{3!\sqrt L (2\pi )^4 }}\sum\limits_{n,n'} {\int\limits_{ - \pi }^\pi  {dx} } \int\limits_{ - \pi }^\pi  {dx'} \int\limits_{ - \pi }^\pi  {dy} \int\limits_{ - \pi }^\pi  {dy'g^{(3)} F^{(3)} (x,x',y,y')} 
\nonumber \\
\left\{ {b_n (x,y)b_{n'} (x',y')} \right.b_{ - n - n'} ( - x - x', - y - y') + 3\gamma (x,y)\hat J_n (x,y)b_{n'} (x,y)b_{ - n - n'} ( - x - x', - y - y')
\nonumber \\
 + 3\gamma (x,y)\gamma (x',y')\hat J_n (x,y)\hat J_{n'} (x',y')b_{ - n - n'} ( - x - x', - y - y') + \gamma (x,y)\gamma (x',y')\gamma ( - x - x', - y - y')
\nonumber \\
\hat J_n (x,y)\hat J_{n'} (x',y')\hat J_{ - n - n'} ( - x - x', - y - y'),
\nonumber
\end{eqnarray}
where
\begin{eqnarray}
F^{(3)} (x,x',y,y') = (1 - \exp ( - ix))(1 - \exp ( - ix'))(1 - \exp (i(x + x')))
\\
+ (1 - \exp ( - iy'))(1 - \exp ( - iy'))(1 - \exp (i(y + y'))),
\nonumber
\end{eqnarray}
\begin{eqnarray}
F_1^{(4)} (x,x',x',y,y',y') = (1 - \exp ( - ix))(1 - \exp ( - ix'))(1 - \exp ( - ix''))(1 - \exp (i(x + x' + x'')))
\nonumber \\
 + (1 - \exp ( - iy))(1 - \exp ( - iy'))(1 - \exp ( - iy''))(1 - \exp (i(y + y' + y''))),
\nonumber
\end{eqnarray}
\begin{eqnarray}
F_2^{(4)} (x,x',x',y,y',y') = (1 - \exp ( - i(x - y)))(1 - \exp ( - i(x' - y')))(1 - \exp ( - i(x'' - y'')))
\nonumber \\
(1 - \exp (i(x - y + x' - y' + x'' - y''))),
\nonumber
\end{eqnarray}
$ g^{(3)}  = a_1 \sin \psi  - 8a_2 \sin 2\psi $, $g_1^{(4)}  = a_1 \cos \psi  - 16a_2 \cos 2\psi$ and
$g_2^{(4)}  = \tilde a_1  - 16\tilde a_2$. Let us calculate the leading order correction to the Gaussian result for 
\begin{eqnarray}
G_n^F (x,y) = \exp \left( { - \frac{{E_0^{AF} }}{{k_B T}}} \right)\frac{1}{Z}\prod\limits_n {\int {\cal D} b_n (x,y)\int {\cal D} \gamma (x,y)} {\rm  }b_n (x,y)b_{ - n} ( - x, - y)
\nonumber \\
\exp \left( { - \frac{{E_H^{AF} [\phi ] + E_L^{AF} [\phi ] + E_{AH}^{AF} [\phi ]}}{{k_B T}}} \right),
\end{eqnarray}
when we neglect $ E_{AH}^{AF} [\phi ]$ and set $E_L^{AF} [\phi ] = 0$, in the Gaussian approximation, we find that $G_n^F (x,y) = k_B TG_n (x,y)$. We derive the leading order correction by expanding out $
E_{AH}^{(1)} [\phi ]$
\begin{eqnarray}
G_n^F (x,y) = k_B TG_n (x,y) + \exp \left( { - \frac{{E_0^{AF} }}{{k_B T}}} \right)\frac{1}{Z}\prod\limits_n {\int {\cal D} b_n (x,y)\int {\cal D} \gamma (x,y)} {\rm  }b_n (x,y)b_{ - n} ( - x, - y)
\nonumber \\
\frac{{E_{AH}^{(1)} [\phi ]}}{{k_B T}}\exp \left( { - \frac{{E_H^{AF} [\phi ]}}{{k_B T}}} \right).
\end{eqnarray}
$E_{AH}^{(2)} [\phi ]$ and $E_L^{AF} [\phi ]$ do not contribute directly to the correction, but do contribute to $\psi$. Using standard procedures we may do the path integrations and so obtain the following result
\begin{equation}
G_n^F (x,y) = k_B TG_n (x,y) + (k_B T)^2 G_n (x,y)^2 (\Delta _p (x,y) + \Delta _f (x,y)),
\end{equation}
where 
\begin{equation}
\Delta _p  = \frac{1}{{2(2\pi )^2 L}}\sum\limits_n {\int\limits_{ - \pi }^\pi  {dx'\int\limits_{ - \pi }^\pi  {dy'} G_n (x',y')\left( {g_1^{(4)} F_1^{(4)} (x, - x,x',y, - y,y') + g_2^{(4)} F_2^{(4)} (x, - x,x',y, - y,y')} \right)} }, 
\end{equation}
and
\begin{eqnarray}
\Delta _f  = \frac{1}{{2(2\pi )^2 L(k_B T)}}\sum\limits_n {\int\limits_{ - \pi }^\pi  {dx'\int\limits_{ - \pi }^\pi  {dy'} S(x',y')\left( {g_1^{(4)} F_1^{(4)} (x, - x,x',y, - y,y') + g_2^{(4)} F_2^{(4)} (x, - x,x',y, - y,y')} \right)} } 
\nonumber \\
\hat J_{ - n'} ( - x', - y')\hat J_{n'} (x',y').
\end{eqnarray}

The first term represents the correction that comes purely from periodic boundary conditions and second term is the contribution from allowing both ends of the molecule to fluctuate independently of each other. Both these corrections  may be represented diagrammatically in Fig. 6. 
\\

\begin{figure}
\includegraphics[12cm,22cm][13cm,23cm]{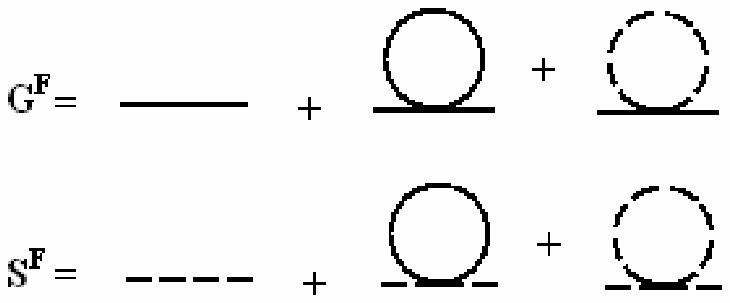}
\vspace{2.5cm} \caption{Diagrammatic representations that contain the 
leading order corrections  to the Gaussian results for the full correlation 
functions $G_n^F(x,y)$ and $S^F(x,y)$ defined in the text.  For $G_n^F(x,y)$  
the first graph from the left 
represents the Gaussian result, the middle graph the correction 
from $\Delta_p$ , 
and the last graph the correction from $\Delta_f$. 
Similarly, for $S^F(x,y)$ the first 
graph from the left represents the Gaussian result, the middle graph the 
correction from $\Sigma_p$ and the last graph the correction from 
$\Sigma_f$.
}
\end{figure}

We may perform the sums which then gives us
\begin{equation}
\Delta _p  = \frac{{\lambda _1 }}{{2C}}\left[ {g_1^{(4)} \chi _1 \left( {\frac{L}{{2\lambda _1^H }},\alpha _H } \right)\left[ {(2 - 2\cos (x)) + (2 - 2\cos (y))} \right] + g_2^{(4)} \chi _2 \left( {\frac{L}{{2\lambda _1^H }},\alpha _H } \right)(2 - 2\cos (x - y)} \right],
\end{equation}
\begin{eqnarray}
\Delta _f  = \frac{{\lambda _1 }}{{(2\pi )^2 C}}\int\limits_{ - L/2}^{L/2} {dx'} \int\limits_{ - L/2}^{L/2} {dy'} \left\{ {g_1^{(4)} \left[ {(2 - 2\cos (x)) + (2 - 2\cos (y))} \right](1 - \cos (x'))} \right.
\nonumber \\
\left. {\left. { + g_2^{(4)} \left[ {(2 - 2\cos (x - y)} \right](1 - \cos (x' - y'))} \right]} \right\}\left( {\frac{{\lambda _1 }}{{2L}}\frac{1}{{\hat C(x',y';\alpha )}}} \right.
\nonumber \\
\left. { - \frac{1}{{4\hat C(x',y';\alpha )^{1/2} }}\left( {\sinh \left( {\frac{{L\hat C(x',y';\alpha )^{1/2} }}{{2\lambda _1 }}} \right)\cosh \left( {\frac{{L\hat C(x',y';\alpha )^{1/2} }}{{2\lambda _1 }}} \right)} \right)^{ - 1} } \right).
\nonumber 
\end{eqnarray}
The functions $\chi _1 \left( {\frac{L}{{2\lambda _1^H }},\alpha _H } \right)$ and 
$ \chi _2 \left( {\frac{L}{{2\lambda _1^H }},\alpha _H } \right) $ are given by
\begin{eqnarray}
\chi _1 \left( {\frac{L}{{2\lambda _1^H }},\alpha _H } \right) = \frac{1}{{(2\pi )^2 }}\int\limits_{ - \pi }^\pi  {dx\int\limits_{ - \pi }^\pi  {dy} } \coth \left( {\frac{L}{{2\lambda _1^H }}\hat C(x,y;\alpha )^{1/2} } \right)\frac{{(1 - \cos x)}}{{\hat C(x,y;\alpha )^{1/2} }},
\nonumber \\
\chi _2 \left( {\frac{L}{{2\lambda _1^H }},\alpha _H } \right) = \frac{1}{{(2\pi )^2 }}\int\limits_{ - \pi }^\pi  {dx\int\limits_{ - \pi }^\pi  {dy} } \coth \left( {\frac{L}{{2\lambda _1^H }}\hat C(x,y;\alpha )^{1/2} } \right)\frac{{(1 - \cos (x - y))}}{{\hat C(x,y;\alpha )^{1/2} }}.
\end{eqnarray}
We may also calculate the leading order correction to the Gaussian result for 
\begin{eqnarray}
S^F (x,y) = \exp \left( { - \frac{{E_0^{AF} }}{{k_B T}}} \right)\frac{1}{Z}\prod\limits_n {\int {\cal D} b_n (x,y)\int {\cal D} \gamma (x,y)} {\rm  }\gamma (x,y)\gamma ( - x, - y)
\nonumber \\
\exp \left( { - \frac{{E_H^{AF} [\phi ] + E_L^{AF} [\phi ] + E_{AH}^{AF} [\phi ]}}{{k_B T}}} \right).
\end{eqnarray}
We find that 
\begin{equation}
S^F (x,y) = S(x,y) + S(x,y)^2 (\Sigma _p (x,y) + \Sigma _f (x,y)),
\end{equation}
where
\begin{eqnarray}
\Sigma _p  = \frac{1}{{2(2\pi )^2 L}}\sum\limits_{n,n'} {\int\limits_{ - \pi }^\pi  {dx'} \int\limits_{ - \pi }^\pi  {dy'} } G_n (x',y')\left( {g_1^{(4)} F_1 (x, - x,x',y, - y,y') + g_2^{(4)} F_2 (x, - x,x',y, - y,y')} \right)
\nonumber \\
\hat J_{n'} (x,y)\hat J_{ - n'} (x,y),
\end{eqnarray}
\begin{eqnarray}
\Sigma _f  = \frac{1}{{2(2\pi )^2 (k_B T)L}}\sum\limits_{n,n'} {\int\limits_{ - \pi }^\pi  {dx'} \int\limits_{ - \pi }^\pi  {dy'} } S(x',y')\left( {g_1^{(4)} F_1 (x, - x,x',y, - y,y') + g_2^{(4)} F_2 (x, - x,x',y, - y,y')} \right)
\nonumber \\
\hat J_n (x',y')\hat J_{n'} (x',y')\hat J_{n''} (x,y)\hat J_{ - n - n' - n''} (x,y). \nonumber
\end{eqnarray}
In Fig. 6, we show how the corrections are represented diagrammatically.
On evaluation of all sums we find
\begin{eqnarray}
\Sigma _f  = \frac{{\lambda _1^2 }}{{2(2\pi )^2 C}} 
{\int\limits_{ - \pi }^\pi  {dx'} 
\int\limits_{ - \pi }^\pi  {dy'} } \left( {g_1^{(4)} ((2 - 2\cos (x)) + (2 - 2\cos (y)))(1 - \cos (x'))} \right.
\nonumber \\
\left. { + g_2^{(4)} (2 - 2\cos (x - y))(1 - \cos (x' - y'))} \right)H\left( {x,y,x',y';\frac{{\lambda _1 }}{L},\alpha } \right),
\end{eqnarray}
\begin{eqnarray}
\Sigma _p  = \left( {\frac{{\lambda _1 ^2 }}{{8C}}} \right)\left[ {\frac{1}{{\hat C(x,y;\alpha )^{1/2} }}\coth \left( {\frac{L}{{2\lambda _1 }}\hat C(x,y;\alpha )^{1/2} } \right) - \frac{L}{{2\lambda _1 }}\left( {\sinh \left( {\frac{L}{{2\lambda _1 }}\hat C(x,y;\alpha )^{1/2} } \right)} \right)^{ - 2} } \right]
\nonumber \\
\frac{1}{{(2\pi )^2 }}\int\limits_{ - \pi }^\pi  {dx'} \int\limits_{ - \pi }^\pi  {dy'} \left( {g_1^{(4)} ((2 - 2\cos (x)) + (2 - 2\cos (y)))(1 - \cos (x')) + g_2^{(4)} (2 - 2\cos (x - y))(1 - \cos (x' - y'))} \right)
\nonumber \\
\frac{1}{{\hat C(x',y';\alpha )^{1/2} }}\coth \left( 
{\frac{{L\hat C(x',y';\alpha )^{1/2} }}{{2\lambda _1 }}} \right), \nonumber
\end{eqnarray}
where
\begin{eqnarray}
H\left( {x,y,x',y';\frac{{\lambda _1 }}{L},\alpha } \right) = \frac{1}{{\hat C(x',y';\alpha )^{1/2} }}\left[ {\frac{1}{4}\coth \left( {\frac{{L\hat C(x,y;\alpha )^{1/2} }}{{2\lambda _1 }}} \right)\left[ {\frac{1}{{\hat C(x',y';\alpha ) - \hat C(x,y;\alpha )}}} \right.} \right.
\nonumber \\
\left. {\left( {\hat C(x',y';\alpha )^{1/2} \coth \left( {\frac{{L\hat C(x,y;\alpha )^{1/2} }}{{\lambda _1 }}} \right) - \hat C(x,y;\alpha )^{1/2} \coth \left( {\frac{{L\hat C(x',y';\alpha )^{1/2} }}{{\lambda _1 }}} \right)} \right)} \right]
\nonumber \\
 + \frac{L}{{8\lambda _1 }}\left( {\sinh \left( {\frac{{L\hat C(x,y;\alpha )^{1/2} }}{{2\lambda _1 }}} \right)} \right)^{ - 2} \left[ {\frac{1}{2}\left( {\sinh \left( {\frac{{L\hat C(x',y';\alpha )^{1/2} }}{{2\lambda _1 }}} \right)\cosh \left( {\frac{{L\hat C(x',y';\alpha )^{1/2} }}{{2\lambda _1 }}} \right)} \right)^{ - 1}  - \frac{{\lambda _1 }}{{L\hat C(x,y;\alpha )^{1/2} }}} \right]
\nonumber \\
 - \frac{1}{{C(x,y;\alpha )^{1/2} 8}}\coth \left( {\frac{{L\hat C(x,y;\alpha )^{1/2} }}{{2\lambda _1 }}} \right)\left. {\left( {\sinh \left( {\frac{{L\hat C(x',y';\alpha )^{1/2} }}{{2\lambda _1 }}} \right)\cosh \left( {\frac{{L\hat C(x',y';\alpha )^{1/2} }}{{2\lambda _1 }}} \right)} \right)^{ - 1} } \right].
\end{eqnarray}
Now let us  determine $\psi$. We do this through the requirement that
\begin{eqnarray}
0 = \frac{1}{N}\sum\limits_{jl} {\left\langle {\phi _{jl} (z) - \phi _{jl - 1} (z)} \right\rangle  = \frac{1}{{(2\pi )^2 \sqrt L }}\sum\limits_n {\exp \left( {\frac{{i\pi nz}}{L}} \right)} } 
\nonumber \\
\left( {\mathop {\mathop {\lim }\limits_{x \to 0} }\limits_{y \to 0} } \right.\left\{ {(1 - \exp ( - iy))\left\langle {b_n (x,y)} \right\rangle } \right\} + \left. {\mathop {\lim }\limits_{\scriptstyle x \to 0 \hfill \atop 
  \scriptstyle y \to 0 \hfill} \left\{ {(1 - \exp ( - iy))J_n \left\langle {\gamma (x,y)} \right\rangle } \right\}} \right).
\end{eqnarray}
Again, in the Gaussian approximation we may neglect $E_{AH}^{AF} [\phi ]$. Equation (E18) is then satisfied by requiring that 
\begin{equation} 
0 = \Gamma _1  =  - {\mathop {\lim }\limits_{\scriptstyle x \to 0 \hfill \atop 
  \scriptstyle y \to 0 \hfill}} \left\{ {(2 - 2\cos y)G_0 (x,y)} \right\}\left( {a_1 \sin \psi _0  - 2a_2 \sin 2\psi _0 } \right),
\end{equation}
which implies that $\psi_0=0$ or $ \psi _0  = \arccos \left( {{{a_1 } \mathord{\left/
 {\vphantom {{a_1 } {(4a_2 )}}} \right.  \kern-\nulldelimiterspace} {(4a_2 )}}} \right)$, where $\psi_0$ is $\psi$ for the Gaussian approximation, for 
$\Gamma _1$ ti vanish.\\

When we go beyond the Gaussian approximation to leading order in perturbation theory we find that $\Gamma_1$ becomes
\begin{eqnarray}
\Gamma _1^F  =  - {\mathop {\lim }\limits_{\scriptstyle x \to 0 \hfill \atop 
  \scriptstyle y \to 0 \hfill}} \left\{ {(2 - 2\cos y)G_0 (x,y)} \right\}\left( {a_1 \sin \psi  - 2a_2 \sin 2\psi } \right)
\nonumber \\
 - {\mathop {\lim }\limits_{\scriptstyle x \to 0 \hfill \atop 
  \scriptstyle y \to 0 \hfill}} \left\{ {(2 - 2\cos y)G_0 (x,y)} \right\}\left( {a_1 \sin \psi  - 8a_2 \sin 2\psi } \right)\left( {\Gamma '_p  + \Gamma '_f } \right),
\end{eqnarray} 
where 
\begin{eqnarray}
\Gamma '_p  = \frac{{k_B T}}{{2(2\pi )^2 L}}\sum\limits_n {\int\limits_{ - \pi }^\pi  {dx'\int\limits_{ - \pi }^\pi  {dy'} } } (2 - 2\cos y')G_n (x',y'),
\nonumber \\
\Gamma '_f  = \frac{1}{{2(2\pi )^2 L}}\sum\limits_n {\int\limits_{ - \pi }^\pi  {dx'\int\limits_{ - \pi }^\pi  {dy'} } } S(x',y')(2 - 2\cos (y'))\hat J_n (x',y')\hat J_{ - n} ( - x', - y').
\end{eqnarray}
Again, the first term, $\Gamma '_p$, arises purely from periodic boundary conditions and $\Gamma '_f$ 
is the contribution from allowing the ends to fluctuate independently of each other. Both these terms may be represented diagrammatically in Fig.7. 
\\

\begin{figure}
\includegraphics[12cm,22cm][13cm,23cm]{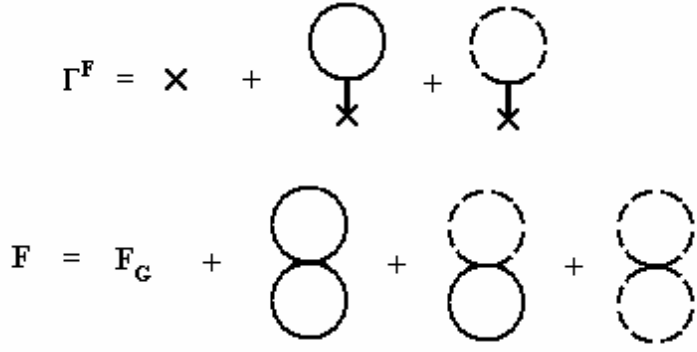}
\vspace{4cm} \caption{Diagrammatic  representations that contain the 
leading order corrections  to the Gaussian results for $\Gamma_1^F(x,y)$   
and the free energy $F$  defined in the text.  For $\Gamma_1^F(x,y)$  
the first graph from the left represents the Gaussian result, the middle 
graph the correction  from $\Delta_p$, and the last graph the correction 
from $\Delta_f$.  For $F$ the first graph from the left represents the 
Gaussian result.. The next graph along is $G_1$, that arises purely from 
periodic boundary conditions. Third from the left is $G_2$ and the last is  
$G_3$.   These terms are corrections due to free boundary conditions.
}
\end{figure}

We may compute the leading order correction to $\psi$ setting $\Gamma_1=0$ and by writing $\psi=\psi_0+\psi'$ and expanding out for small $\psi'$. We may also perform the summations in both  $\Gamma '_p$ and $\Gamma '_f$. We find that
\begin{equation}
\psi ' =  - \frac{{(a_1 \sin \psi _0  - 8a_2 \sin 2\psi _0 )}}{{(a_1 \cos \psi _0  - 4a_2 \cos 2\psi _0 )}}\left( {\Gamma '_p  + \Gamma '_f } \right),
\end{equation}
where now
\begin{eqnarray}
\Gamma '_p  = \frac{{\lambda _1 }}{{4\lambda _p }}\chi _1 \left( {\frac{L}{{2\lambda _1^H }},\alpha _H } \right), \nonumber
 \\
\Gamma '_f  = \frac{{\lambda _1^2 }}{{(2\pi )^2 4\lambda _p L}} {\int\limits_{ - \pi }^\pi  {dx'\int\limits_{ - \pi }^\pi  {dy'} } } (1 - \cos (y'))\left[ {\frac{1}{{\hat C(x',y';\alpha )}}} \right. - \frac{L}{{2\lambda _1 }}\frac{1}{{\hat C(x',y';\alpha )^{1/2} }}
\\
\left. {\left( {\sinh \left( {\frac{{L\hat C(x',y';\alpha )^{1/2} }}{{2\lambda _1 }}} \right)\cosh \left( {\frac{{L\hat C(x',y';\alpha )^{1/2} }}{{2\lambda _1 }}} \right)} \right)^{ - 1} } \right]. \nonumber
\end{eqnarray}
\\

 We may also look at corrections to the free energy. We may write $F=F_G-k_B T \Delta F$ where $F_G$ is 
$F$ evaluated in the Gaussian approximation (c.f (C10)) and  $\Delta F$   is the correction. We find for
$\Delta F$
\begin{equation}
\Delta F = G_1  + G_2  + G_3, 
\end{equation}
\begin{eqnarray}
G_1  = \frac{1}{8}\frac{{k_B T}}{{L(2\pi )^4 }}\sum\limits_{n,n'} {\int\limits_{ - \pi }^\pi  {dx} \int\limits_{ - \pi }^\pi  {dy} \int\limits_{ - \pi }^\pi  {dx'} \int\limits_{ - \pi }^\pi  {dy'} } G_n (x,y)G_{n'} (x',y')
\nonumber \\
\left. {\left( {g_1^{(4)} F_1^{(4)} (x, - x,x',y, - y,y')} \right) + g_2^{(4)} F_2^{(4)} (x, - x,x',y, - y,y')} \right), 
\nonumber
\end{eqnarray}
\begin{eqnarray}
G_2  = \frac{1}{4}\frac{1}{{L(2\pi )^4 }}\sum\limits_{n,n'} {\int\limits_{ - \pi }^\pi  {dx} \int\limits_{ - \pi }^\pi  {dy} \int\limits_{ - \pi }^\pi  {dx'} \int\limits_{ - \pi }^\pi  {dy'} } S(x,y)G_{n'} (x',y')\hat J_n (x,y)\hat J_{ - n} (x,y)
\nonumber \\
\left. {\left( {g_1^{(4)} F_1^{(4)} (x, - x,x',y, - y,y')} \right. + g_2^{(4)} F_2^{(4)} (x, - x,x',y, - y,y')} \right),
\nonumber
\end{eqnarray}
\begin{eqnarray}
G_3  = \frac{1}{8}\frac{1}{{k_B TL(2\pi )^4 }}\sum\limits_{n,n',n''} {\int\limits_{ - \pi }^\pi  {dx} \int\limits_{ - \pi }^\pi  {dy} \int\limits_{ - \pi }^\pi  {dx'} \int\limits_{ - \pi }^\pi  {dy'} } S(x,y)S(x',y')\hat J_n (x,y)\hat J_{n'} (x,y)\hat J_{n''} (x',y')
\nonumber \\
\hat J_{ - n - n' - n''} (x',y')\left( {g_1^{(4)} F_1^{(4)} (x, - x,x',y, - y,y') + g_2^{(4)} F_2^{(4)} (x, - x,x',y, - y,y')} \right). \nonumber
\end{eqnarray}
These corrections are represented graphically in Fig. 7.  The sums may, again, be evaluated yielding the following expressions
\begin{equation}
G_1  = \frac{L}{8}\frac{{k_B T}}{{(2\pi )^4 }}\left( {\frac{{\lambda _1 }}{C}} \right)^2 \left[ {2g_1^{(4)} \chi _1 \left( {\frac{L}{{2\lambda _1^H }},\alpha _H } \right)^2  + g_2^{(4)} \chi _2 \left( {\frac{L}{{2\lambda _1^H }},\alpha _H } \right)^2 } \right],
\end{equation}
\begin{eqnarray}
G_2  = \frac{L}{2}\frac{{k_B T}}{{(2\pi )^4 }}\left( 
{\frac{{\lambda _1 }}{C}} \right)^2 \int\limits_{ - \pi }^\pi  
{dx\int\limits_{ - \pi }^\pi  {dy} \int\limits_{ - \pi }^\pi  {dx'} } 
\int\limits_{ - \pi }^\pi  {dy'} \left( {2g_1^{(4)} (1 - \cos x)(1 - \cos x')} \right.
\nonumber \\
\left. { + g_2^{(4)} (1 - \cos (x - y))(1 - \cos (x' - y'))} \right)\frac{1}{{\hat C(x,y;\alpha )^{1/2} }}\coth \left( {\frac{{L\hat C(x,y;\alpha )^{1/2} }}{{2\lambda _1 }}} \right)
\nonumber \\
\left[ {\frac{{\lambda _1 }}{{2L}}\frac{1}{{\hat C(x',y';\alpha )}} - \frac{1}{4}\frac{1}{{\hat C(x',y';\alpha )^{1/2} }}\left( {\sinh \left( {\frac{{LC(x',y';\alpha )^{1/2} }}{{2\lambda _1 }}} \right)\cosh \left( {\frac{{LC(x',y';\alpha )^{1/2} }}{{2\lambda _1 }}} \right)} \right)^{ - 1} } \right],
\nonumber
\end{eqnarray}
\begin{eqnarray}
G_3  = \frac{1}{4}\frac{{\lambda _1^2 }}{{C(2\pi )^4 }}\int\limits_{ - \pi }^\pi  {dx} \int\limits_{ - \pi }^\pi  {dy} \int\limits_{ - \pi }^\pi  {dx'} \int\limits_{ - \pi }^\pi  {dy'} \left[ {2g_1^{(4)} (1 - \cos x)(1 - \cos x')} \right.
\nonumber \\
\left. { + g_2^{(4)} (1 - \cos (x - y))(1 - \cos (x' - y'))} \right]S(x,y)H\left( {x,y,x',y';\frac{L}{{\lambda _1 }},\alpha } \right). \nonumber
\end{eqnarray}
In $G_n^F (x,y)$, $S^F(x,y)$ and $F$ there are additional corrections that come from inserting $\psi$ into $G_n(x,y)$ and $S(x,y$ and $F_G$; then expanding these terms out to leading order in $\psi'$.
For the purposes of illustration we will show 
\begin{equation}
G_n^F (x,y) = k_B TG_n (x,y) + (k_B T)^2 \left[ {\Delta _f^H  + [(2 - 2\cos (x) + (2 - 2\cos (y)]\psi '(a_1 \sin \psi _0  - 8a_2 \sin 2\psi _0 )/(k_B T)} \right]G_n (x,y)^2. 
\end{equation}
\\

\renewcommand{\theequation}{F\arabic{equation}}
\setcounter{equation}{0}
\section*{Appendix F. Derivation of self consistent approximation for periodic boundary conditions.}
Now, for the moment, let us concentrate on  periodic boundary conditions and set $\Gamma(x,y)=0$. 
We may represent each term in the perturbation theory diagrammatically. The Feynman rules; the correspondence between a graph and its algebraic expression are now given. On expanding out $
E[\phi (\vec k)]$ we need three types of diagrams (or vertices) 
to represent each $\phi^n$-term in this expansion, for $n>0$. The Feynman 
rules are similar to those given in Appendix A of \cite{wynveen:05a}, but with a few modifications \cite{fixfeyn}. They are as follows.  Each graph will contain $N_F$ vertices all of which will be connected to each other by lines. We assign a label $i = 1 \ldots N_V$ to each vertex. For each vertex $i$, representing $\phi^n$, we must write down  the following for a type 1 or 2 vertex 
\begin{eqnarray}
\left( {a_1 ( - 1)^{n/2 - 1} \cos (\psi ) + a_2 ( - 4)^{n/2} \cos (2\psi )} \right)\delta _{ - k^i _1 ,k^i _{2n}  + k^i _{2n - 1}  +  \ldots k^i _2 } /n!
\,\, {\rm when } \,\, n \,\, {\rm is}\,\,{\rm even}
\nonumber \\
\left( {a_1 ( - 1)^{(n + 1)/2 - 1} \sin (\psi ) + a_2 \frac{{( - 4)^{(n + 1)/2} }}{2}\sin (2\psi )} \right)\delta _{ - k^i _1 ,k^i _{2n}  + k^i _{2n - 1}  +  \ldots k^i _2 } /n!
\,\,{\rm when } \,\, n \,\, {\rm is}\,\,{\rm odd} \nonumber \\  
\end{eqnarray}
and either one of the ``form'' factors 
\begin{equation}
\prod\limits_{m = 1}^n {\left( {1 - e^{ - ix_m^i } } \right)} ,{\rm  }\prod\limits_{m = 1}^n {\left( {1 - e^{ - iy_m^i } } \right)} ,\,\,\,
\end{equation}
depending on whether the vertex is type $1$  or $2$ ,  respectively. For a type $3$ vertex, $n$ is allowed only to be even, and we write down
\begin{equation}
\left( {\tilde a_1 ( - 1)^{n/2 - 1}  + \tilde a_2 ( - 4)^{n/2} } \right)\delta _{ - k^i _1 ,k^i _{2n}  + k^i _{2n - 1}  +  \ldots k^i _2 } /n!,
\end{equation}
 multiplied by the form factor 
\begin{equation}
\prod\limits_{m = 1}^n {\left( {1 - e^{ - i(x_m^i  - y_m^i )} } \right)}. 
\end{equation}
\\

Each full Feynman graph will also consist of $N_E$ external 
lines (connected to only one vertex) each associated with a wave vector $\vec q_j  = \left( {x_j /r_0 ,y_j /r_0 ,2\pi n_j /L} \right)$ $(j = 1, \ldots ,N_E )$. For each of these external lines we write down $G_{n_i } (x_i ,y_i )$ and set $\vec k_m^i  = \vec q_j$ for one of the $\vec k_m^i  = \left( {x_m^i /r_0 ,y_m^i /r_0 ,2\pi n_m^i /L} \right)
$ in the vertex to which the line is connected. There will also 
be $N_I$ internal lines, each associated with a wave vector $\vec p_k$$(k = 1, \ldots ,N_I )$, where each end is connected to two vertices, $i$ and $i'$. For each of these internal lines we write down $G(\vec p_k )$ and set $\vec k_m^i  = \vec k_m^{i'}  = \vec p_j$ for one of the $\vec k_m^i$ in each of the two vertices. Then all the wave vectors for the internal lines are summed over. Last of all, there is also a symmetry factor that multiplies this, which accounts for how many ways a term (graph) in the expansion may be generated. If we restrict the wave vectors to the form $\vec q_j  = \left( {x_j /r_0 ,y_j /r_0 ,0} \right)$ we get back the results for the rigid body case \cite{wynveen:05a}. 
\\

To obtain the Hartree approximation we first consider the same set of  graphs as for Appendix A of \cite{wynveen:05a} as contributions to the full correlation function.   The sum of these graphs we denote by $G_{1n}(x,y)$.
These form a series which we may easily sum  
\begin{eqnarray}
\frac{{k_B TG_{1n} ^{ - 1} (x,y)}}{C} = \left( {\frac{{2\pi n}}{L}} \right)^2  + \left( {\lambda _{1,1} \left( {\frac{L}{{\lambda _1 }},\alpha ,\frac{{\lambda _1 }}{{\lambda _p }}} \right)} \right)^{ - 2} \left( {(1 - \cos x) + (1 - \cos y)} \right)
\nonumber \\
 + \left( {\lambda _{2,1} \left( {\frac{L}{{\lambda _1 }},\alpha ,\frac{{\lambda _1 }}{{\lambda _p }}} \right)} \right)^{ - 2} (1 - \cos (x - y)),
\end{eqnarray}
where
\begin{eqnarray} 
\lambda _{1,1} \left( {\frac{L}{{\lambda _1 }},\alpha ,\frac{{\lambda _1 }}{{\lambda _p }}} \right) = \sqrt {\frac{C}{{2\left( {a_1 \cos (\psi )\exp \left( { - \frac{{\lambda _1 \chi _1 \left( {L/(2\lambda _1 ),\alpha } \right)}}{{4\lambda _p }}} \right) - 4a_2 \cos (2\psi )\exp \left( { - \frac{{\lambda _1 \chi _1 \left( {L/(2\lambda _1 ),\alpha } \right)}}{{\lambda _p }}} \right)} \right)}}}, 
\nonumber \\
\lambda _{2,1} \left( {\frac{L}{{\lambda _1 }},\alpha ,\frac{{\lambda _1 }}{{\lambda _p }}} \right) = \sqrt {\frac{C}{{2\left( {\tilde a_1 \exp \left( { - \frac{{\lambda _1 \chi _2 \left( {L/(2\lambda _1 ),\alpha } \right)}}{{4\lambda _p }}} \right) - 4\tilde a_2 \exp \left( { - \frac{{\lambda _1 \chi _2 \left( {L/(2\lambda _1 ),\alpha } \right)}}{{\lambda _p }}} \right)} \right)}}}. 
\end{eqnarray}

In the same procedure as discussed in Appendix A of \cite{wynveen:05a}, 
we may  replace $G_n$ in each loop (of our diagrammatic series) 
with $G_{1n}$. Now. $G_{1n}$ will be replaced on the l.h.s. of (F5) 
with a new correlation function $G_{2n}$. On the r.h.s of (F5)  
we replace $\lambda_{1,1}$ with $\lambda_{1,2}$
and $\lambda_{2,1}$ with $\lambda_{2,2}$. Expressions 
for $\lambda_{1,2}$ and  $\lambda_{2,2}$ are similar  to (F6), 
but with $\lambda$ replaced by $\lambda_{1,1}$ and $\alpha$ replaced by 
$\alpha _1  = \left( {\lambda _{1,1} /\lambda _{2,1} } \right)^2$. 
We then keep iterating this process until we have $ \lambda _1^H  = \lambda _{1,\infty }  = \lambda _{1,\infty  - 1} $, $\lambda _2^H  = \lambda _{2,\infty }  = \lambda _{2,\infty  - 1}$ and $\alpha _H  = \left( {\lambda _1^H /\lambda _2^H } \right)^2$ so obtain Eq. (4.3) of the text. Through same reasoning as was discussed in Appendix. B of \cite{wynveen:05a},  we find that $\psi=\psi_H$ where
\begin{equation}
\cos \psi _H  = \frac{{a_1 }}{{4a_2 }}\exp \left( {\frac{{3\lambda _1 \chi _1 \left( {L/(2\lambda _1 ),\alpha } \right)}}{{4\lambda _p }}} \right).
\end{equation}
\\

To calculate the free energy we consider the same set of graphs as those 
considered for free energy in \cite{wynveen:05a} .To get the free energy 
in the Hartree approximation we then renormalize the sum of these graphs, 
by taking care in replacing $\lambda_1$ with $\lambda _1^H$, $\lambda_2$ 
with $\lambda _2^H$, and $\alpha$ with $\alpha_H$; and so arrive at  
Eq. (4.2) of the text.

%%%%%%%%%%%%%%%%%%%%%%%%%%%%%%%%%%%%%%%%%%%%%%%%%%%%%%%%
\renewcommand{\theequation}{G\arabic{equation}}
\setcounter{equation}{0}
\section*{Appendix G.  Correction to self consistent approximation for periodic boundary conditions from freely fluctuating ends.}
To calculate the corrections arising from free fluctuating ends we must 
utilize the results of both the previous two appendices. We perform a 
renormalization where we replace $\lambda_1$, $\lambda_2$, $\alpha$ and 
$\psi_0$ with $\lambda _1^H$, $\lambda _2^H$, $\alpha_H$ and $\psi_H$ in 
the results of Appendix E. The renormalization process relies on 
counter-terms which take care of terms already included in $\lambda _1^H$, 
$\lambda _2^H$, $\alpha_H$ and $\psi_H$, to prevent over-counting. In the 
interests of brevity we will not discuss this process of renormalization, 
instead we refer those not acquainted with such processes to a standard 
text in field theoretical methods.
\\

 The first quantity to consider is $\psi'$ which on  renormalization becomes
\begin{equation}
\psi ' _H  =  - \frac{{(a_1 \sin \psi _H  - 8a_2 \sin 2\psi _H )}}
{{(a_1 \cos \psi _H  - 4a_2 \cos 2\psi _H )}}\left( 
{\Gamma _f ^{'H} } \right),
\end{equation}
where
\begin{eqnarray}
\Gamma _f ^{'H}  = \frac{{\left( {\lambda _1^H } \right)^2 }}{{4(2\pi )^2 
\lambda _p L}} {\int\limits_{ - \pi }^\pi  
{dx'\int\limits_{ - \pi }^\pi  {dy'} } } (1 - \cos (y'))\left[ {\frac{1}{{\hat C(x',y';\alpha _H )}}} \right. - \frac{L}{{2\lambda _1 }}\frac{1}{{\hat C(x',y';\alpha _H )^{1/2} }}
\nonumber \\
\left. {\left( {\sinh \left( {\frac{{L\hat C(x',y';\alpha _H )^{1/2} }}{{2\lambda _1^H }}} \right)\cosh \left( {\frac{{L\hat C(x',y';\alpha _H )^{1/2} }}{{2\lambda _1^H }}} \right)} \right)^{ - 1} } \right].
\end{eqnarray}
It is important to notice that there is no $\Gamma _p ^{'H}$, this is because such a term is already accounted for in $\psi_H$. On renormalization, a counter term removes this term.  
\\

Let us next consider $G_n^F (x,y)$, to obtain the correction from free boundary conditions we perform  a renormalization of (E8) where we obtain
\begin{eqnarray}
G_n^{F,R} (x,y) = k_B TG_n^H (x,y) + (k_B T)^2 G_n^H (x,y)^2 
\nonumber \\
\left( {\Delta _f^H  + [(2 - 2\cos (x) + (2 - 2\cos (y)]\psi '_H (a_1 \sin \psi _H  - 8a_2 \sin 2\psi _H )/(k_B T)} \right),
\end{eqnarray}
where
\begin{equation}
\Delta _f^H  =  {\left[ {(2 - 2\cos (x)) + (2 - 2\cos (y))} \right]} \Delta _f^{1,H}  + (2 - 2\cos (x - y))\Delta _f^{2,H} ,
\end{equation}
\begin{eqnarray}
\Delta _f^{1,H}  = \frac{{\lambda _1^H g_1^{(4)} }}{{(2\pi )^2 C}}\int\limits_{ - L/2}^{L/2} {dx'} \int\limits_{ - L/2}^{L/2} {dy'} (1 - \cos x')\left( {\frac{{\lambda {}_1^H }}{{2L}}\frac{1}{{\hat C(x',y';\alpha _H )}}} \right.
\nonumber \\
 - \frac{1}{{4\hat C(x',y';\alpha _H )^{1/2} }}\left. {\left( {\sinh \left( {\frac{{L\hat C(x',y';\alpha _H )^{1/2} }}{{2\lambda _1^H }}} \right)\cosh \left( {\frac{{L\hat C(x',y';\alpha _H )^{1/2} }}{{2\lambda _1^H }}} \right)} \right)^{ - 1} } \right),
\nonumber
\end{eqnarray}
\begin{eqnarray}
\Delta _f^{2,H}  = \frac{{\lambda _1^H g_2^{(4)} }}{{(2\pi )^2 C}}\int\limits_{ - L/2}^{L/2} {dx'} \int\limits_{ - L/2}^{L/2} {dy'} (1 - \cos (x' - y'))\left( {\frac{{\lambda {}_1^H }}{{2L}}\frac{1}{{\hat C(x',y';\alpha _H )}}} \right.
\nonumber \\
 - \frac{1}{{4\hat C(x',y';\alpha _H )^{1/2} }}\left. {\left( {\sinh \left( {\frac{{L\hat C(x',y';\alpha _H )^{1/2} }}{{2\lambda _1^H }}} \right)\cosh \left( {\frac{{L\hat C(x',y';\alpha _H )^{1/2} }}{{2\lambda _1^H }}} \right)} \right)^{ - 1} } \right),
\nonumber
\end{eqnarray}
and
\begin{eqnarray}
\frac{{k_B TG_n^H (x,y)^{ - 1} }}{C} = \left( {\frac{{2\pi n}}{L}} \right)^2  + \left( {\lambda _1^H } \right)^{ - 2} \left( {(1 - \cos x) + (1 - \cos y)} \right)
\nonumber \\
 + \left( {\lambda _2^H } \right)^{ - 2} (1 - \cos (x - y)).
\end{eqnarray}
There is  no $\Delta _p^H$, this is because such a term is already accounted for in $G_n^H (x,y)$, so is removed  on renormalization. We may go further and make the approximation that
\begin{eqnarray}
G_n^{F,R} (x,y) \simeq k_B TG_n^H (x,y) + (k_B T)^2 G_n^H (x,y)^2 \left( {\Delta _f^H  + [(2 - 2\cos (x) + (2 - 2\cos (y)]\psi '_H } \right.
\nonumber \\
\left. {(a_1 \sin \psi _H  - 8a_2 \sin 2\psi _H )/(k_B T)} \right) + (k_B T)^3 G_n^H (x,y)^3 \left( {\Delta _f^H  + [(2 - 2\cos (x) + (2 - 2\cos (y))]} \right.
\nonumber \\
\left. {\psi '_H (a_1 \sin \psi _H  - 8a_2 \sin 2\psi _H )/(k_B T)} \right)^2  +  \ldots 
\nonumber \\
 \simeq k_B TG_n^H (x,y) + (k_B T)^2 \left( {\Delta _f^H  + [(2 - 2\cos (x) + (2 - 2\cos (y)]\psi '_H (a_1 \sin \psi _H  - 8a_2 \sin 2\psi _H )/(k_B T)} \right)
\nonumber \\
G_n^H (x,y)G_n^{F,R} (x,y).
\end{eqnarray}
From (G6) we may write
\begin{eqnarray}
\frac{{k_B TG_n^{F,R} (x,y)^{ - 1} }}{C} = \left( {\frac{{2\pi n}}{L}} \right)^2  + \left( {\lambda _1^F } \right)^{ - 2} \left( {(1 - \cos x) + (1 - \cos y)} \right)
\nonumber \\
 + \left( {\lambda _2^F } \right)^{ - 2} (1 - \cos (x - y)),
\end{eqnarray}
where
\begin{eqnarray}
\left( {\lambda _1^F } \right)^{ - 2}  = \left( {\lambda _1^H } \right)^{ - 2}
  - 2C^{ - 1} \left( {k_B T\Delta _f^{1,H}  + \psi '_H (a_1 \sin \psi _H  - 8a_2 \sin 2\psi _H )} \right)
\nonumber \\
\left( {\lambda _2^F } \right)^{ - 2}  = 
\left( {\lambda _2^H } \right)^{ - 2}  - 2C^{ - 1} k_B T\Delta _f^{2,H}. 
\end{eqnarray}
Now, we renormalize $S^F(x,y)$ 
\begin{eqnarray}
S^{F,R} (x,y) = S^H (x,y) + S^H (x,y)^2 \left( {\Sigma _f^H  - } \right.\frac{{(a_1 \sin \psi _H  - 8a_2 \sin 2\psi _H )\psi '_H \lambda _1^H }}{{2k_B T}}[(1 - \cos x) + (1 - \cos y)]
\nonumber \\
\left. {\left( {\frac{L}{{2\lambda _1^R }}\left( {\sinh \left( {\frac{{L\hat C(x,y;\alpha )^{1/2} }}{{2\lambda _1^R }}} \right)} \right)^{ - 2}  - \frac{1}{{\hat C(x,y;\alpha )}}\coth \left( {\frac{{L\hat C(x,y;\alpha )^{1/2} }}{{2\lambda _1^R }}} \right)} \right)} \right),
\end{eqnarray}
\begin{eqnarray}
\Sigma _f^H  = \frac{{\left( {\lambda _1^H } \right)^2 }}{{2(2\pi )^2 C}}
 {\int\limits_{ - \pi }^\pi  {dx'} \int\limits_{ - \pi }^\pi  {dy'} } \left( {g_1^{(4)} ((2 - 2\cos (x)) + (2 - 2\cos (y)))(1 - \cos (x'))} \right.
\nonumber \\
\left. { + g_2^{(4)} (2 - 2\cos (x - y))(1 - \cos (x' - y'))} \right)H\left( {x,y,x',y';\frac{{\lambda _1^H }}{L},\alpha _H } \right).
\nonumber
\end{eqnarray}
There is  no $\Sigma _p^H$ because such a term is already accounted for in $S^F(x,y)$.  \\

We may also renormalize the free energy, the free energy then becomes 
$F_T  = F_H  + F_F  + \Delta F$ where 
 \begin{eqnarray}
F_F  = \frac{{k_B T}}{{2(2\pi )^2 }}\int\limits_{ - \pi }^\pi  {dx} \int\limits_{ - \pi }^\pi  {dy\ln \left( {\frac{{\lambda _p }}{{\lambda _1^H }}\left( {\frac{{\hat C(x,y;\alpha _H )}}{{\hat C(x,y;1)}}} \right)^{1/2} \coth \left( {\frac{{\hat C(x,y;\alpha _H )^{1/2} L}}{{2\lambda _1^H }}} \right)} \right)} ,\nonumber \\
\Delta F =  - L(\psi '_H )^2 m_1  - k_B TG_3^H,
\end{eqnarray} 
and
\begin{eqnarray}
G_3^H  = \frac{1}{4}\frac{{\left( {\lambda _1^H } \right)^2 }}{{C(2\pi )^4 }}\int\limits_{ - \pi }^\pi  {dx} \int\limits_{ - \pi }^\pi  {dy} \int\limits_{ - \pi }^\pi  {dx'} \int\limits_{ - \pi }^\pi  {dy'} \left[ {2g_1^{(4)} (1 - \cos x)(1 - \cos x')} \right.
\nonumber \\
\left. { + g_2^{(4)} (1 - \cos (x - y))(1 - \cos (x' - y'))} \right]S(x,y)H\left( {x,y,x',y';\frac{L}{{\lambda _1^H }},\alpha _H } \right).
\end{eqnarray}
We may also calculate correlation functions within this approximation scheme. Here, we may use the results of the previous section, but we now make the following replacements: $\lambda_1$ is replaced by
$\lambda _1^F$, $\alpha$ is replaced by $\alpha _F  = (\lambda _1^F /\lambda _2^F )^2$ and $S(x,y)$ is 
replaced by $S^{F,R} (x,y)$.

\renewcommand{\theequation}{H\arabic{equation}}
\setcounter{equation}{0}
\section*{Appendix H.  High Temperature Expansion.}
We start the analysis of this appendix by dividing  (3.1) into three terms
\begin{equation}
E[\phi ] = E_{{\mathop{\rm int}} }^0  + E'_{{\mathop{\rm int}} } [\phi ] + E_T [\phi ],
\end{equation}
where
\begin{equation}
E_{{\mathop{\rm int}} }^0  = 3a_0 (R_1 ),
\end{equation}
\begin{eqnarray}
E'_{{\mathop{\rm int}} } [\phi ] = \int\limits_{ - L/2}^{L/2} {dz} \sum\limits_{j,l} {\sum\limits_{m=1}^{\infty} {\left[ {( - 1)^m a_m (R_1 )\left( {\cos \left( {m\left( {\phi _{jl} (z) - \phi _{j + 1l - 1} (z)} \right)} \right)} \right.} \right.} } 
\nonumber \\
\left. { + \cos \left( {m\left( {\phi _{jl} (z) - \phi _{j - 1l} (z)} \right)} \right) + \cos \left( {m\left( {\phi _{jl} (z) - \phi _{jl - 1} (z)} \right)} \right)} \right],
\nonumber \\
E_T [\phi ] = \int\limits_{ - L/2}^{L/2} {dz} \sum\limits_{j,l} {\frac{C}{2}\left( {\frac{{d\phi _{jl} (z)}}{{dz}}} \right)^2 } .
\nonumber 
\end{eqnarray}
First, let us look at the high temperature expansion of the free energy. We start by  expanding out the partition function in the following way
\begin{equation}
Z = \exp \left( { - \frac{{E_{{\mathop{\rm int}} }^0 }}{{k_B T}}} \right)\prod\limits_{jl} {\int {\cal D} \phi (z)} {\rm  }\left( {1 - \frac{{E'_{{\mathop{\rm int}} } [\phi ]}}{{k_B T}} + \frac{1}{2}\left( {\frac{{E'_{{\mathop{\rm int}} } [\phi ]}}{{k_B T}}} \right)^2  +  \ldots } \right)\exp \left( { - \frac{{E_T [\phi ]}}{{k_B T}}} \right).
\end{equation}
From this expansion the lowest order terms in $Z$ that depend on the coefficients $a_n$ are
\begin{eqnarray}
Z_1^m  = \frac{{a_m^2 }}{{2(k_B T)^2 }}\exp \left( { - \frac{{E_{{\mathop{\rm int}} }^0 }}{{k_B T}}} \right)\prod\limits_{j,l} {\int {\cal D} \phi _{jl}^p (z)} \int {d} \gamma _{jl} \int {d} \phi _{jl}^0 \sum\limits_{j',l'} {\sum\limits_{j'',l''} {\int\limits_{ - L/2}^{L/2} {dz\int\limits_{ - L/2}^{L/2} {dz'} } } } 
\nonumber \\
\left[ {\cos \left( {m\left( {\tilde \phi _{j'l'}^p (z) - \tilde \phi _{j' - 1l'}^p (z) + \frac{z}{L}\left( {\gamma _{j'l'}  - \gamma _{j' - 1l'} } \right) + \phi _{j'l'}^0  - \phi _{j' - 1l'}^0 } \right)} \right)} \right. + 
\nonumber \\
\cos \left( {m\left( {\tilde \phi _{j'l'}^p (z) - \tilde \phi _{j'l' - 1}^p (z) + \frac{z}{L}\left( {\gamma _{j'l'}  - \gamma _{j'l' - 1} } \right) + \phi _{j'l'}^0  - \phi _{j'l' - 1}^0 } \right)} \right) + 
\nonumber \\
\left. {\cos \left( {m\left( {\tilde \phi _{j'l'}^p (z) - \tilde \phi _{j' + 1l' - 1}^p (z) + \frac{z}{L}\left( {\gamma _{j'l'}  - \gamma _{j' + 1l' - 1} } \right) + \phi _{j'l'}^0  - \phi _{j' + 1l' - 1}^0 } \right)} \right)} \right]
\nonumber \\
\left[ {\cos \left( {m\left( {\tilde \phi _{j''l''}^p (z') - \tilde \phi _{j'' - 1l''}^p (z') + \frac{{z'}}{L}\left( {\gamma _{j''l''}  - \gamma _{j'' - 1l''} } \right) + \phi _{j''l''}^0  - \phi _{j'' - 1l''}^0 } \right)} \right)}
 \right. \nonumber \\
 + \cos \left( {m\left( {\tilde \phi _{j''l''}^p (z') - \tilde \phi _{j''l'' - 1}^p (z') + \frac{{z'}}{L}\left( {\gamma _{j''l''}  - \gamma _{j''l'' - 1} } \right) + \phi _{j''l''}^0  - \phi _{j''l'' - 1}^0 } \right)} \right) + 
\nonumber \\
\left. {\cos \left( {m\left( {\tilde \phi _{j''l''}^p (z') - \tilde \phi _{j'' + 1l'' - 1}^p (z') + \frac{{z'}}{L}\left( {\gamma _{j''l''}  - \gamma _{j'' + 1l'' - 1} } \right) + \phi _{j''l''}^0  - \phi _{j'' + 1l'' - 1}^0 } \right)} \right)} \right]\exp \left( { - \frac{{E_T [\phi ]}}{{k_B T}}} \right),
\end{eqnarray}
where we have employed (3.7) of the text but, now separating out the 
rigid body mode $\phi _{jl}^0$ so that $\phi _{jl}^p (z) = \tilde \phi _{jl}^p
 (z) + \phi _{jl}^0$, and the spatial average of $\tilde \phi _{jl}^p (z)$ 
along the length of the molecule is zero.  We find that the only a few of these terms  survive integration over the rigid body modes.
\begin{eqnarray}
Z_1^m  = \frac{{a_m^2 (2\pi )^N }}{{4(k_B T)^2 }}\exp \left( { - \frac{{E_{{\mathop{\rm int}} }^0 }}{{k_B T}}} \right)\prod\limits_{j,l} {\int {\cal D} \phi _{jl}^p (z)} \int {d} \gamma _{jl} \sum\limits_{j',l'} {\int\limits_{ - L/2}^{L/2} {dz\int\limits_{ - L/2}^{L/2} {dz'} } } \exp \left( { - \frac{{E_T [\phi ]}}{{k_B T}}} \right)
\nonumber \\
\left[ {\cos \left( {m\left( {\tilde \phi _{j'l'}^p (z) - \tilde \phi _{j'l'}^p
 (z') - \tilde \phi _{j' - 1l'}^p (z) + \tilde \phi _{j' - 1l'}^p (z') + \left( {z - z'} \right)\left( {\gamma _{j'l'}  - \gamma _{j' - 1l'} } \right)/L} \right)} \right)} \right. + 
\nonumber \\
\cos \left( {m\left( {\tilde \phi _{j'l'}^p (z) - \tilde \phi _{j'l'}^p (z')} \right.} \right. - \tilde \phi _{j'l' - 1}^p (z) + \tilde \phi _{j'l' - 1}^p (z')\left. {\left. { + \left( {z - z'} \right)\left( {\gamma _{j'l'}  - \gamma _{j'l' - 1} } \right)/L} \right)} \right)
\nonumber \\
 + \cos \left( {m\left( {\tilde \phi _{j'l'}^p (z) - \tilde \phi _{j'l'}^p (z') - \tilde \phi _{j' + 1l' - 1}^p (z) + \tilde \phi _{j' + 1l' - 1}^p (z')} \right.} \right.\left. {\left. {\left. { + (z - z')\left( {\gamma _{j'l'}  - \gamma _{j' + 1l' - 1} } \right)/L} \right)} \right)} \right].
\end{eqnarray}
Now we may rewrite $ E_T [\phi ]$ as $E_T  = E_T^f  + E_T^p$ where
\begin{equation}
E_T^f  = \sum\limits_{j,l} {\frac{C}{{2L}}\gamma _{jl}^2 } 
\,\, {\rm and}\,\, 
E_T^p  = \sum\limits_{j,l} {\frac{C}{2}\left( {\frac{{d\tilde \phi _{jl}^p }}{{dz}}} \right)} ^2 .
\end{equation}
We find that we may then write the following expansion of $\ln Z$ 
\begin{equation}
\ln Z \simeq \ln Z_0  + \sum\limits_m {Z_1^m /Z_0 }  +  \ldots ,
\end{equation}
\begin{eqnarray}
&&\frac{{Z_1^m }}{{Z_0 }} = \frac{{a_n^2 }}{{4(k_B T)^2 }}\sum\limits_{j',l'} {\int\limits_{ - L/2}^{L/2} {dz\int\limits_{ - L/2}^{L/2} {dz'} } } \exp \left( { - \frac{{E_T [\phi ]}}{{k_B T}}} \right)
\nonumber \\
&& \times \left[ {\left\langle {\cos \left( {m\left( {\tilde \phi _{j'l'}^p (z) - \tilde \phi _{j'l'}^p (z') - \tilde \phi _{j' - 1l'}^p (z) + \tilde \phi _{j' - 1l'}^p (z')} \right)} \right)} \right\rangle _{E_T^p } \left\langle {\cos \left( {m\left( {\left( {z - z'} \right)\left( {\gamma _{j'l'}  - \gamma _{j' - 1l'} } \right)/L} \right)} \right)} \right\rangle } \right._{E_T^f } 
\nonumber \\
&& + \left\langle {\cos \left( {m\left( {\tilde \phi _{j'l'}^p (z) - \tilde \phi _{j'l'}^p (z')} \right.} \right. - \tilde \phi _{j'l' - 1}^p (z)\left. {\left. { + \tilde \phi _{j'l' - 1}^p (z')} \right)} \right)} \right\rangle _{E_T^p } \left\langle {\cos \left( {m\left( {\left( {z - z'} \right)\left( {\gamma _{j'l'}  - \gamma _{j'l' - 1} } \right)/L} \right)} \right)} \right\rangle _{E_T^f } 
\nonumber \\
&& + \left\langle {\cos \left( {m\left( {\tilde \phi _{j'l'}^p (z) - \tilde \phi _{j'l'}^p (z') - \tilde \phi _{j' + 1l' - 1}^p (z) + \tilde \phi _{j' + 1l' - 1}^p (z')} \right)} \right)} \right\rangle _{E_T^p } \left. {\left\langle {\cos \left( {m\left( {(z - z')\left( {\gamma _{j'l'}  - \gamma _{j' + 1l' - 1} } \right)/L} \right)} \right)} \right\rangle _{E_T^f } } \right] \nonumber
\end{eqnarray}
where
\begin{eqnarray}
&& \left\langle {\cos \left( {m\left( {\tilde \phi _{j'l'}^p (z) - \tilde \phi _{j'l'}^p (z') - \tilde \phi _{j' - 1l'}^p (z) + \tilde \phi _{j' - 1l'}^p (z')} \right)} \right)} \right\rangle _{E_T^p } 
\nonumber \\
&& = \frac{{\prod\limits_{j,l} {\int {\cal D} \phi _{jl}^p (z)} \cos \left( {m\left( {\tilde \phi _{j'l'}^p (z) - \tilde \phi _{j'l'}^p (z') - \tilde \phi _{j' - 1l'}^p (z) + \tilde \phi _{j' - 1l'}^p (z')} \right)} \right)\exp \left( { - \frac{{E_T^p [\phi ]}}{{k_B T}}} \right)}}{{\prod\limits_{j,l} {\int {\cal D} \phi _{jl}^p (z)\exp \left( { - \frac{{E_T^p [\phi ]}}{{k_B T}}} \right)} }},
\end{eqnarray}
\begin{equation}
\left\langle {\cos \left( {m\left( {\frac{{\left( {z - z'} \right)}}{L}\left( {\gamma _{j'l'}  - \gamma _{j' - 1l'} } \right)} \right)} \right)} \right\rangle _{E_T^p }  = \frac{{\prod\limits_{j,l} {\int {d\gamma _{jl} } } \cos \left( {m\left( {\frac{{\left( {z - z'} \right)}}{L}\left( {\gamma _{j'l'}  - \gamma _{j' - 1l'} } \right)} \right)} \right)\exp \left( { - \frac{{E_T^f [\phi ]}}{{k_B T}}} \right)}}{{\prod\limits_{j,l} {\int {d\gamma _{jl} } \exp \left( { - \frac{{E_T^f [\phi ]}}{{k_B T}}} \right)} }} \nonumber
\end{equation}
The term $\left\langle {\cos \left( {n\left( {\left( {z - z'} \right)\left( {\gamma _{j'l'}  - \gamma _{j' - 1l'} } \right)/L} \right)} \right)} \right\rangle _{E_T^p }$ is easy to evaluate and we find
\begin{eqnarray}
\left\langle {\cos \left( {m\left( {\frac{{\left( {z - z'} \right)}}{L}\left( {\gamma _{j'l'}  - \gamma _{j' - 1l'} } \right)} \right)} \right)} \right\rangle _{E_T^p }  = \left\langle {\cos \left( {m\left( {\frac{{\left( {z - z'} \right)}}{L}\left( {\gamma _{j'l'}  - \gamma _{j'l' - 1} } \right)} \right)} \right)} \right\rangle _{E_T^p } 
\nonumber \\
 = \left\langle {\cos \left( {m\left( {\frac{{\left( {z - z'} \right)}}{L}\left( {\gamma _{j'l'}  - \gamma _{j' + 1l' - 1} } \right)} \right)} \right)} \right\rangle _{E_T^p }  = \exp \left( { - \frac{{m^2 \left( {z - z'} \right)^2 }}{{2\lambda _p L}}} \right).
\end{eqnarray}
The term $\left\langle {\cos \left( {m\left( {\tilde \phi _{j'l'}^p (z) 
- \tilde \phi _{j'l'}^p (z') - \tilde \phi _{j' - 1l'}^p (z) 
+ \tilde \phi _{j' - 1l'}^p (z')} \right)} \right)} \right\rangle _{E_T^p } $ 
is slightly more involved, where to evaluate this term we 
express $\tilde \phi _{j'l'}^p (z)$ in terms of its Fourier series 
$\tilde \phi _{jl}^p (z) = \sum\limits_{n \ne 0} {b_{jl}^n \exp 
\left( {\frac{{2i\pi n z}}{L}} \right)}$ we then find that
\begin{eqnarray}
\left\langle {\cos \left( {m\left( {\tilde \phi _{j'l'}^p (z) 
- \tilde \phi _{j'l'}^p (z') - \tilde \phi _{j' - 1l'}^p (z) 
+ \tilde \phi _{j' - 1l'}^p (z')} \right)} \right)} \right\rangle _{E_T^p } 
\nonumber \\
\left\langle {\cos \left( {m\left( {\tilde \phi _{j'l'}^p (z) - 
\tilde \phi _{j'l'}^p (z') - \tilde \phi _{j'l' - 1}^p (z) + 
\tilde \phi _{j'l'-1}^p (z')} \right)} \right)} \right\rangle _{E_T^p } 
\nonumber \\
 = \left\langle {\cos \left( {m\left( {\tilde \phi _{j'l'}^p (z) - 
\tilde \phi _{j'l'}^p (z') - \tilde \phi _{j' + 1l' - 1}^p (z) 
+ \tilde \phi _{j' + 1l' - 1}^p (z')} \right)} \right)} 
\right\rangle _{E_T^p } 
\nonumber \\
 = \exp \left( { - \sum\limits_n {\frac{{2Lm^2 }}{{\lambda _p (2\pi n)^2 }}} \left( {1 - \cos \left( {\frac{{2\pi n(z - z')}}{L}} \right)} \right)} \right).
\end{eqnarray}
The sum may be evaluated leaving us with
\begin{equation}
\left\langle {\cos \left( {m\left( {\tilde \phi _{j'l'}^p (z) 
- \tilde \phi _{j'l'}^p (z') - \tilde \phi _{j' + 1l' - 1}^p (z) + \tilde \phi _{j' + 1l' - 1}^p (z)} \right)} \right)} \right\rangle _{E_T^p }  = \exp \left( { - \frac{{m^2 \left| {z - z'} \right|}}{{2\lambda _p }}} \right)\exp \left( {\frac{{m^2 \left| {z - z'} \right|^2 }}{{2\lambda _p L}}} \right).
\end{equation}
Putting this all together we find
\begin{equation}
\frac{{Z_1^m }}{{Z_0 }} = \frac{{3a_m^2 N}}{{4(k_B T)^2 }}
\int\limits_{ - L/2}^{L/2} {dz\int\limits_{ - L/2}^{L/2} {dz'} } 
\exp \left( { - \frac{{\left| {z - z'} \right|}}{{2\lambda _p }}} 
\right) = \frac{{3a_m^2 L^2 N}}{{4(k_B T)^2 }}
f_2\left( {\frac{{m^2 L}}{{\lambda _p }}} \right),
\end{equation}
where
\begin{equation}
f_2(x) = \frac{{[8x + 8\exp ( - x/2) - 8]}}{{x^2 }}.
\end{equation}
If we truncate the sum in (H1) over the $a_m$, according to 
\cite{kornyshev:97a} for a DNA molecule,  at $m=2$, then we may also calculate the next to leading order correction to the free energy per molecule, 
\begin{eqnarray}
F = F_0  - \frac{{3(a_1^2 L^2 f_2 (L/\lambda _p ) + a_2^2 L^2 f_2 (4L/\lambda _p ))}}{{4(k_B T)}} - \frac{{(a_1^3 L^3 f_{3,1} (L/\lambda _p ) - a_2^3 L^3 f_{3,1} (4L/\lambda _p ))}}{{2(k_B T)^2 }}
\nonumber \\
+ \frac{{3a_2 a_1^2 L^3 f_{3,2} (L/\lambda _p )}}{{8(k_B T)^2 }} +  \ldots ,
\end{eqnarray}
where $F_0$ is a term independent of $a_1$ and $a_2$, and
\begin{eqnarray}
f_{3,1} (x) = \frac{{24[\exp ( - x/2)(4 + x) + ( - 4 + x)]}}{{x^3 }},
\nonumber \\
f_{3,2} (x) = \frac{{[\exp ( - x/2)(128 + 24x) + 36x - 126 - 2\exp ( - 2x)]}}{{3x^3 }}.
\end{eqnarray}\\

We may look at limiting cases. When the molecules are either very short or rigid, so that $ L \ll \lambda _p$, then we find that we may write
\begin{equation}
F = F_0  - \frac{{3L^2 (a_1^2  + a_2^2 )}}{{4(k_B T)}} - \frac{{(a_1^3 L^3  - a_2^3 L^3 )}}{{2(k_B T)^2 }} + \frac{{3a_2 a_1^2 L^3 }}{{8(k_B T)^2 }} +  \ldots .
\end{equation}
This yields the following specific heat
\begin{equation}
\frac{{C_v }}{{k_B }} = \frac{3}{2}\left( {\frac{{(La_1 )^2  + (La_2 )^2 }}{{k_B T}}} \right) + 3\left( {\frac{{(La_1 )^3  - (La_2 )^3 }}{{\left( {k_B T} \right)^2 }}} \right) - \frac{{9a_2 a_1^2 L^3 }}{{4\left( {k_B T} \right)^2 }} +  \ldots .
\end{equation}
Here we have an additional term when compared with [20], this term was overlooked previously. We may also look at the limit where molecules are very long or soft. Here we find
\begin{equation}
F = F_0  - \frac{{3(4a_1^2 L\lambda _p  + a_2^2 L\lambda _p )}}{{2(k_B T)}} - \frac{{(48a_1^3 L\lambda _p^2  - 3a_2^3 L\lambda _p^2 )}}{{4(k_B T)^2 }} + \frac{{9a_2 a_1^2 L\lambda _p^2 }}{{2(k_B T)^2 }} +  \ldots .
\end{equation}
When comparing the two limits we see that as we move from short rigid molecules to long flexible molecules our high temperature expansion  changes from a power series expansions in $La_1 /(k_B T)$ and
$La_2 /(k_B T)$ to  power series expansions in $\lambda _p a_1 /(k_B T)$ and $\lambda _p a_2 /(k_B T)$.
So for $ L \gg \lambda _p $, the high temperature expansion being valid when we have $\lambda _p a_1 /(k_B T) \ll 1$ and $\lambda _p a_2 /(k_B T) \ll 1$, not necessarily $L a_1 /(k_B T) \ll 1$ and $L a_2 /(k_B T) \ll 1$. 
It is also interesting to look at the correlation function      
\begin{equation}
\left\langle {\exp \left( {in\left( {\phi _{jl} (z) - \phi _{j'l'} (z')} \right)} \right)} \right\rangle  = {\rm  }\frac{1}{Z}\prod\limits_{jl} {\int {\cal D} \phi (z)} \exp \left( {in\left( {\phi _{jl} (z) - \phi _{j'l'} (z')} \right)} \right)\exp \left( { - \frac{{E[\phi ]}}{{k_B T}}} \right)
\end{equation}
evaluated for nearest and next nearest neighbors. To generate the high temperature expansion  we make the following expansion 
\begin{eqnarray}  
\left\langle {\exp \left( {in\left( {\phi _{jl} (z) - \phi _{j'l'} (z')} \right)} \right)} \right\rangle  = \left( {\prod\limits_{jl} {\int {\cal D} \phi (z)} {\rm  }\left( {1 - \frac{{E'_{{\mathop{\rm int}} } [\phi ]}}{{k_B T}} + \frac{1}{2}\left( {\frac{{E'_{{\mathop{\rm int}} } [\phi ]}}{{k_B T}}} \right)^2  +  \ldots } \right)\exp \left( { - \frac{{E_T [\phi ]}}{{k_B T}}} \right)} \right)^{ - 1} 
\nonumber \\
\left( {\prod\limits_{jl} {\int {\cal D} \phi (z)} {\rm  }\left( {1 - \frac{{E'_{{\mathop{\rm int}} } [\phi ]}}{{k_B T}} + \frac{1}{2}\left( {\frac{{E'_{{\mathop{\rm int}} } [\phi ]}}{{k_B T}}} \right)^2  +  \ldots } \right)\exp \left( {in\left( {\phi _{jl} (z) - \phi _{j'l'} (z')} \right)} \right)\exp \left( { - \frac{{E_T [\phi ]}}{{k_B T}}} \right)} \right)
\end{eqnarray}
In general the leading order term in the correlation function- that describes how azimuthal correlations are lost- is of the form
\begin{equation}
\left\langle {\exp \left( {in\left( {\phi _{jl} (z) - \phi _{j'l'} (z')} 
\right)} \right)} \right\rangle  \simeq n_b \left( {\frac{{a_n }}{{2k_B T}}} 
\right)^b \tilde f_b \left( 
{\frac{z}{L},\frac{{z'}}{L};\frac{{n^2 L}}{{\lambda _p }}} \right)
\end{equation}
Here, $b$ is the smallest number of links or bonds between sites $(i,j)$ and $(i',j')$. A link being defined as a translation of $ \pm R_1 {\bf \hat u}$, $ \pm R_1 {\bf \hat v}$ or $\pm R_1 \left( {{\bf \hat v} - {\bf \hat u}} \right)$ between two sites on the 2-D lattice. The factor $n_b$ is the number of paths using the smallest number of links that go between the two sites $(i,j)$ and $(i',j')$. This is illustrated for the examples that we give explicit results in Fig. 8. Also the scaling function should always have the property that for torsionally rigid molecules $ \tilde f_b  = 1$. 
\\

\begin{figure}
\includegraphics[12cm,22cm][13cm,23cm]{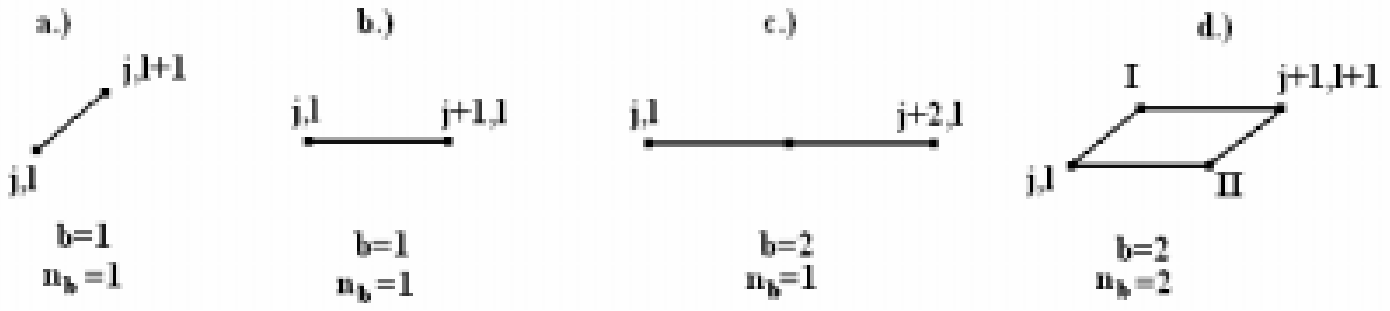}
\vspace{4cm} \caption{The paths with the shortest number of links for the 
correlation functions a.) $\left\langle \exp \left( in \left(\phi_{jl}(z)-
\phi_{jl+1}(z') \right) \right) \right\rangle$    
b.)   $\left\langle \exp \left( in \left(\phi_{jl}(z)-
\phi_{j+1l}(z') \right) \right) \right\rangle$    
c.) $\left\langle \exp \left( in \left(\phi_{jl}(z)-
\phi_{j+2l}(z') \right) \right) \right\rangle$       and  
d.)  $\left\langle \exp \left( in \left(\phi_{jl}(z)-
\phi_{j+1l+1}(z') \right) \right) \right\rangle$. Now, 
a.) and 
b.) contain only one link, while 
c.) and 
d.) 
two. d.) 
is the only one, shown here,  to have  two different paths, path I and II.
}
\end{figure}

First let us calculate the correlation function where $b=1$. This is the correlation function between nearest neighbors. We find to leading order in the high temperature expansion. 
 \begin{eqnarray}
\left\langle {\exp \left( {in\left( {\phi _{jl} (z) - \phi _{j - 1l} (z')} \right)} \right)} \right\rangle  = \left\langle {\exp \left( {in\left( {\phi _{jl} (z) - \phi _{jl - 1} (z')} \right)} \right)} \right\rangle 
\nonumber \\
 = \left\langle {\exp \left( {in\left( {\phi _{jl} (z) - \phi _{j + 1l - 1} (z')} \right)} \right)} \right\rangle  = \frac{{a_n ( - 1)^n }}{{2k_B T}}\tilde f_1 \left( {\frac{z}{L},\frac{{z'}}{L};\frac{{n^2 L}}{{\lambda _p }}} \right),
\end{eqnarray}
where
\begin{eqnarray}
f_1 (x,x';y) = \frac{1}{y}\left( {4\exp \left( { - \frac{{y\left| {x - x'} \right|}}{4}} \right)} \right. + y\left| {x - x'} \right|\exp \left( { - \frac{{y\left| {x - x'} \right|}}{4}} \right)
\nonumber \\
\left. { - 2\exp \left( { - \frac{y}{4}} \right)\left( {\exp \left( {\frac{{y(x + x')}}{4}} \right) + \exp \left( { - \frac{{y(x + x')}}{4}} \right)} \right)} \right).
\end{eqnarray}
When $ \lambda _p  \gg L$ we do, indeed, find that $f_1 (x,x';y) \simeq 1$.
\\

We may also compute correlations where $b=2$. This is between next to nearest neighbors and next to next to nearest order correlations. We find to leading order\begin{eqnarray}
2\left\langle {\exp \left( {in\left( {\phi _{jl} (z) - \phi _{j - 2l} (z')} \right)} \right)} \right\rangle  = 2\left\langle {\exp \left( {in\left( {\phi _{jl} (z) - \phi _{jl - 2} (z')} \right)} \right)} \right\rangle 
\nonumber \\
= 2\left\langle {\exp \left( {in\left( {\phi _{jl} (z) - \phi _{j + 2l - 2} (z')} \right)} \right)} \right\rangle  = \left\langle {\exp \left( {in\left( {\phi _{jl} (z) - \phi _{j - 1l - 1} (z')} \right)} \right)} \right\rangle 
\nonumber \\
 = \left\langle {\exp \left( {in\left( {\phi _{jl} (z) - \phi _{j - 2l + 1} (z')} \right)} \right)} \right\rangle  = \left\langle {\exp \left( {in\left( {\phi _{jl} (z) - \phi _{j - 1l + 2} (z')} \right)} \right)} \right\rangle 
\nonumber \\
 = \frac{{a_n ( - 1)^n }}{{2k_B T}}\tilde f_2 \left( {\frac{z}{L},\frac{{z'}}{L};\frac{{n^2 L}}{{\lambda _p }}} \right),
\end{eqnarray}
where
\begin{eqnarray}
\tilde f_2 \left( {x,x';y} \right) = \frac{{4\tilde f\left( {x,x';y} \right)}}{y} + \frac{8}{{y^2 }}\exp \left( { - \frac{{y\left| {x - x'} \right|}}{4}} \right)
\nonumber \\
 - 2\left( {\frac{1}{y} + 4\frac{1}{{y^2 }}} \right)\exp \left( { - \frac{y}{4}} \right)\left[ {\exp \left( {\frac{{y(x + x')}}{4}} \right) + \exp \left( { - \frac{{y(x + x')}}{4}} \right)} \right]
\nonumber \\
 + \frac{{2(x + x')}}{y}\exp \left( { - \frac{y}{4}} \right)\left[ {\exp \left( {\frac{{y(x + x')}}{4}} \right) - \exp \left( { - \frac{{y(x + x')}}{4}} \right)} \right]
\nonumber \\
 + \left( {\frac{{2\left| {x - x'} \right|}}{y} + \frac{{(x - x')^2 }}{2}} \right)\exp \left( { - \frac{{y\left| {x - x'} \right|}}{4}} \right)
\nonumber \\
 + \frac{4}{{y^2 }}\exp \left( { - \frac{y}{2}} \right)\left[ {\exp \left( {\frac{{y(x - x')}}{4}} \right) + \exp \left( { - \frac{{y(x - x')}}{4}} \right)} \right].
\end{eqnarray}
Again, when $ \lambda _p  \gg L$, we find that $f_2 (x,x';y) \simeq 1$.
\\

We may also look at self correlations ($b=0$), due to torsional fluctuations, for a single DNA molecule
\begin{equation}
\left\langle {\exp \left( {in\left( {\phi _{jl} (z) - \phi _{jl} (z')} \right)} \right)} \right\rangle  = \exp \left( {\frac{{ - n^2 \left| {z - z'} \right|}}{{4\lambda _p }}} \right).
\end{equation}

%%%%%%%%%%%%%%%%%%%%%%%%%%%%%%%%%%%%%%%%%%%%%%%%%%%%%%%%%%%%%%%%%%%%%

%%%%%%%%%%%%%%%%%%%%%%%%%%%%%%%%%%%%%%%%%%%%%%%%%%%%%%%%%%%%%%%%%%%
%%%%%%%%%%%%%%%%%%%%%%%%%%%%%%%%%%%%%%%%%%%%%%%%%%%%%%%%%%%%%%%%%%%
%%%%%%%%%%%%%%%%%%%%%%%%%%%%%%%%%%%%%%%%%%%%%%%%%%%%%%%%%%%%%%%%%%%
%%%%%%%%%%%%%%%%%%%%%%%%%%%%%%%%%%%%%%%%%%%%%%%%%%%%%%%%%%%%%%%%%%%
%%%%%%%%%%%%%%%%%%%%%%%%%%%%%%%%%%%%%%%%%%%%%%%%%%%%%%%%%%%%%%%%%%%
%%%%%%%%%%%%%%%%%%%%%%%%%%%%%%%%%%%%%%%%%%%%%%%%%%%%%%%%%%%%%%%%%%%
%%%%%%%%%%%%%%%%%%%%%%%%%%%%%%%%%%%%%%%%%%%%%%%%%%%%%%%%%%%%%%%%%%%
%%%%%%%%%%%%%%%%%%%%%%%%%%%%%%%%%%%%%%%%%%%%%%%%%%%%%%%%%%%%%%%%%%%
%%%%%%%%%%%%%%%%%%%%%%%%%%%%%%%%%%%%%%%%%%%%%%%%%%%%%%%%%%%%%%%%%%%
%%%%%%%%%%%%%%%%%%%%%%%%%%%%%%%%%%%%%%%%%%%%%%%%%%%%%%%%%%%%%%%%%%%


\bibliography{mybib}

\begin{thebibliography}{36}
\bibitem[{bkz()}]{bkz}
\bibinfo{note}{For the range of parameters considered here, the
  Berezinskii-Kosterlitz-Thouless state would occur at extremely high
  temperatures.}
\bibitem[{flu()}]{flucends}
\bibinfo{note}{Note that Eq. (2.8) is not a unique way of ensuring that $\phi
  (-L/2) + \gamma = \phi (L/2)$. We could have chosen a different ansatz.
  Nevertheless, the end result is ansatz-independent. Any ansatz that satisfies
  this criterion will yield the same form for the partition function as long as
  the correct variable shift to decouple $b_n$ from $\gamma$ is chosen.}
\bibitem[{azd()}]{azde}
\bibinfo{note}{Since our primary concern is to study the effects of azimuthal
  fluctuations of the molecules, we do not consider their spatial fluctuations
  about the lattice points. Though coupling between azimuthal fluctuations and
  spatial fluctuations does occur, this can be neglected within the first
  approximation.}
\bibitem[{fix()}]{fixfeyn}
\bibinfo{note}{There was a mistake in the Feynman rules stated in Ref. [20],
  namely that each vertex represents $\phi{^{2n}}$ and not $\phi{^n}$. However,
  the correct expression for the Hartree result was obtained in Appendix A of
  Ref. [20].}
\end{thebibliography}

\end{document}